# Simulation Modelling and Analysis of Primary Health Centre Operations

Mohd Shoaib and Varun Ramamohan

*Department of Mechanical Engineering, Indian Institute of Technology Delhi, New Delhi 110016, India*

MOHD SHOAIB is a Ph.D. candidate in the Department of Mechanical Engineering at the Indian Institute of Technology Delhi. His email address is mez178322@mech.iitd.ac.in.

VARUN RAMAMOHAN* is an Assistant Professor in the Department of Mechanical Engineering at the Indian Institute of Technology Delhi. He holds a Ph.D. in Industrial Engineering from Purdue University, Indiana, USA. His email address is varunr@mech.iitd.ac.in.

*Corresponding author: Varun Ramamohan
Postal address: III-358, Department of Mechanical Engineering, Indian Institute of Technology Delhi, New Delhi 110016, India.

# Simulation Modelling and Analysis of Primary Health Centre Operations

## Abstract

We present discrete-event simulation models of the operations of primary health centres (PHCs) in the Indian context. Our PHC simulation models incorporate four types of patients seeking medical care: outpatients, inpatients, childbirth cases, and patients seeking antenatal care. A generic modelling approach was adopted to develop simulation models of PHC operations. This involved developing an archetype PHC simulation, which was then adapted to represent two other PHC configurations, differing in numbers of resources and types of services provided, encountered during PHC visits. A model representing a benchmark configuration conforming to government-mandated operational guidelines, with demand estimated from disease burden data and service times closer to international estimates (higher than observed), was also developed. Simulation outcomes for the three observed configurations indicate negligible patient waiting times and low resource utilisation values at observed patient demand estimates. However, simulation outcomes for the benchmark configuration indicated significantly higher resource utilisation. Simulation experiments to evaluate the effect of potential changes in operational patterns on reducing the utilisation of stressed resources for the benchmark case were performed. Our analysis also motivated the development of simple analytical approximations of the average utilisation of a server in a queueing system with characteristics similar to the PHC doctor/patient system. Our study represents the first step in an ongoing effort to establish the computational infrastructure required to analyse public health operations in India, and can provide researchers in other settings with hierarchical health systems a template for the development of simulation models of their primary healthcare facilities.

## 1. Introduction

Providing quality healthcare services in India is a challenge given the rapidly increasing demand due to the aging population, growing health-seeking behaviour among the population due to increased awareness, and spurt in the burden of noncommunicable diseases. This is exacerbated by the inadequate size of the public health workforce.[1] These challenges are more pronounced in rural regions because of socioeconomic factors such as increased poverty,[2,3] illiteracy,[4] and high levels of social inequality.[3] The situation is further exacerbated because nearly seventy percent of the population resides in rural areas,[5] which faces an acute shortage of trained medical staff. According to Jaiswal et al.,[6] nearly seventy percent of doctors in India are based in cities while seventy percent of the demand arises from the villages. Additionally, in rural India as few as thirty-seven percent people have access to inpatient facilities within a 5 kilometre distance, and only sixty-eight percent have access to an outpatient department (OPD).[7] Finally, despite the economic burden of availing private healthcare, only twenty percent of the people seeking outpatient care and forty-five percent seeking inpatient care utilise public healthcare services.[8]

Outside large public hospitals in urban metropolitan areas that provide highly specialised tertiary care (called superspecialty hospitals), the public health system in India comprises three levels of formal medical care: the primary health centre (PHC, which offers primary care), the community health centre (CHC, which offers primary and limited secondary care), and the district hospital (DH, which offers comprehensive secondary and limited tertiary care).[9-11] Further, smaller facilities known as subcentres (SCs) focus on public health awareness and immunisation programmes and are present primarily in rural areas. PHCs form the basic unit of public healthcare delivery in India, and represent the first point of contact for the patient with a formally trained medical doctor. Hence, as part of efforts to address the issues described in the previous paragraph and increase the utilisation of public health services, there is increasing interest in strengthening primary healthcare delivery through establishing new PHCs, upgrading existing primary health infrastructure, and increasing medical personnel numbers. Specifically, the government has recently announced plans to create 150,000 Health and Wellness Centres (HWCs) by financial year 2022.[12] Under this scheme all subcentres and primary health centres will be upgraded to HWCs to deliver universal and free comprehensive primary health care to the public. Although there has been a considerable increase in the number of PHCs across the country, i.e., from 9,115 during 1981-85 to 25,650 in 2017,[13] their operational effectiveness and influence on improving public health accessibility is not adequately quantified. In this backdrop, there is need for an assessment of the operational aspects of these facilities before more resources are invested in their upgradation and/or establishing new PHC infrastructure.

Further, an effort to comprehensively model public health operations in the Indian context would require developing simulation models of PHCs, given their foundational importance in the public health system (PHCs outnumber CHCs by a ratio of 5:1).[14] Therefore, in this study, we focus on developing discrete-event simulation (DES) models of PHC operations as part of an ongoing effort to establish the computational infrastructure required to model and analyse public health operations in the Indian context.[15] Our approach towards modelling and analysing PHC operations can provide researchers and analysts in other countries with similar hierarchical public health systems with a template for developing models of similar primary/secondary healthcare facilities in their contexts. For example, the Ghanaian public health care system consists of five tiers (similar to the Indian system when subcentres and superspecialty hospitals are also considered), with their 'subdistrict' health centres being the equivalent of the Indian PHCs.[16] Similarly, the public health system in Bangladesh is organised into four levels: community level healthcare (provided by the domiciliary health workers and community clinics), primary level healthcare (Union Health and Family Welfare Centres (UH&FWCs), and Upazila Health Complexes), secondary level healthcare (provided in district hospitals, general hospitals, among others), and tertiary level healthcare (provided in postgraduate medical institutes, and other large hospitals). UH&FWCs are the equivalent of Indian PHCs, and house one staff member with formal medical training and support staff that focus on delivering limited curative outpatient care, maternal and child health care.

In public health systems such as in India and the other countries described above, while there may be commonalities in operational patterns of these facilities because they are established according to operational guidelines specified by a central public health planning authority, there may also be significant differences in operational configurations between facilities. A generic modelling approach thus becomes an effective way to model such facilities, as the operational commonalities can be captured by a generic model developed by surveying multiple instances of these facilities. Subsequently, to capture the operational diversity in these facilities that is characteristic of such health systems, the surveyed facilities can be grouped into configuration classes which can then be modelled by adapting/reusing the generic/archetypal simulation model. Thus, a key research contribution of our study is the demonstration of this approach for modelling PHCs, which can, as described above, prove useful for modelling hierarchical public health systems in other settings as well.

Our approach - the generic modelling approach - involved visiting multiple PHCs in a semi-urban/rural district in North India and collecting data regarding their operational patterns. We then develop an archetypal or 'generic' DES model of PHC operations based on the commonalities in PHC operations observed during our visits, and subsequently adapt (reuse with modification) this generic model to represent the different operational configurations encountered in our visits. We then compare the performance of these existing PHC configurations with the performance of a benchmark configuration conforming to government-mandated operational guidelines, with demand estimated from disease burden data and service times closer to international estimates, which are significantly higher than observed service times at the PHCs. Our literature search did not identify any studies that adopted such an approach, driven by public health data and international healthcare delivery practice, towards operational benchmarking of such healthcare facilities. Thus another research contribution of our study involves the demonstration of this approach.

The benchmarking exercise also motivated the conduct of simulation experiments to quantify how these PHC configurations respond to changes (increases) in demand, and identify solutions to potentially improve operational efficiency under conditions of high demand. We anticipate that the model and such analyses can provide key stakeholders with a methodology to make informed decisions regarding changes in PHC/HWC operational guidelines or when upgrading or establishing existing/new PHCs/HWCs.

In the Indian context, to our knowledge, computational studies on the operational patterns and performance of PHCs have not been done before. Our research contribution in this context thus involves the estimation - via DESs - of operational outcomes such as the average waiting time of patients for various resources (e.g., doctors, pharmacy, clinical laboratory), resource utilisation levels across the PHC, and the proportions of childbirth patients who wait longer than a certain time threshold. Note that a DES is not strictly required to obtain rough estimates of average wait times. From the patient load and service time data we collect for the PHC doctors, for example, one can anticipate negligible outpatient wait times. However, exact quantification of these wait times and resource utilisation levels is not straightforward. For example, the outpatient care system consists of two queues in series - certain patients undergo an initial consultation with a nurse followed by consulting with the doctor, and the doctors themselves serve three types of patients with very different arrival rates. Further, the interaction between various queueing subsystems within the PHC - e.g., the outpatient care, the pharmacy, and laboratory subsystems - yield somewhat counterintuitive results under certain conditions, as will be seen in Section 4. Finally, developing PHC simulations represents the first step in establishing the computational infrastructure required to conduct other operational analyses of the public healthcare system (e.g., how would implementation of rigorously enforced referral mechanisms change operational outcomes across the public health network in a district), with similar simulation models of CHCs and DHs to follow.

In the general healthcare delivery simulation context, while there are simulation studies which model patient flow in a single unit such as the outpatient clinic [17,18] the emergency department,[19] or in multiple interdependent units (emergency department and inpatient department (IPD)) serving a single patient type,[20] we found very limited work that utilises a generic modelling approach for simulating primary healthcare delivery facilities that handle multiple patient types (with distinct clinical and operational flows through multiple facility units) and services (the PHC provides outpatient care, childbirth, antenatal services, limited inpatient care, pharmacy and clinical laboratory services). Thus, another key research contribution of this study involves addressing the above research gap. Finally, we also develop two analytical approximations of the utilisation of a server with characteristics similar to that of the PHC doctor (multiple job types, each with Markovian arrival rates and generally distributed service times) that resulted from the internal validation exercises that we conducted for the PHC simulations.

This article is organised as follows. In Section 2, we provide an overview of the relevant literature, and in Section 3, we elaborate on the generic modelling approach adopted in the study, PHC operational data collection, clinical and operational flow in PHCs, and input parameter estimation. In Section 4, we describe model validation efforts, key simulation outcomes, sensitivity and other operational analyses, and we conclude with a discussion of the study in Section 5.

## 2. Background and Literature Review

In this section, we provide a brief overview of PHCs, and then describe the literature related to the application of DES in healthcare facility modelling. A brief overview of the public healthcare infrastructure in India is provided in Appendix A.

In India, PHCs are established to deliver integrated curative, promotive, and preventive healthcare services. They provide OPD services for six days a week and also operate 24 hour emergency services. A PHC is intended to serve a population ranging from 30,000 persons in the plains and 20,000 persons in mountainous or heavily forested areas. Mainly outpatient services are rendered in a typical PHC; however, they house a delivery hut to assist in infant deliveries and have a small inpatient unit for patients requiring care and observation for brief periods. This could include, for instance, management of injuries and accidents, dengue, diarrhoea etc.[10]

A typical PHC may house one or two doctors (decided based on the monthly childbirth load typical to the region), a pharmacist, one laboratory technician, three to four staff nurses working in shifts, and other non-medical support staff. In addition to a delivery hut, a PHC must have four to six beds for catering to inpatients/emergency cases. Apart from these, PHCs are also responsible for community engagement, which is managed through subcentres. Community engagement is driven by auxillary nurse midwives (ANMs) or multipurpose health workers, and involves creating awareness for hygiene and infectious diseases, maternal health, childcare, distribution of essential medicines/supplements, etc. PHCs can also refer patients who require more specialised or intensive care to the CHCs or the district hospital.

## 2.1. Related Work

Simulation has been used in virtually all segments of the healthcare delivery analysis field. Simulation applications in healthcare include modelling for staffing decisions, facility design and location, patient flow, appointment scheduling, capacity allocation, and logistics. Comprehensive literature surveys on simulation applications in healthcare have been published.[21-25] DES is the most widely used simulation methodology, and this is reflected in the publication of a number of survey articles regarding the use of DES in healthcare,[22,26-28] which readers can refer to for a comprehensive account of the relevant literature.

We begin by discussing a relevant review article by Gunal and Pidd [27], who classified the relevant literature based on the healthcare unit modelled: accident and emergency units (i.e., emergency/casualty/trauma units), inpatient facilities, outpatient clinics, other hospital units (intensive care units (ICUs), laboratories, etc.), and whole hospitals. Of particular relevance, the authors discussed the study by Fetter and Thompson,[29] who developed simulation models of hospital subsystems not specific to a facility, and instead described them in general. The subsystems considered by them are: a) maternity department, b) outpatient department, and c) surgical department. These subsystem simulation models were used for evaluating different operating policies and design changes.

There are numerous simulation articles that deal with modelling a specific unit of a healthcare facility such as the outpatient department, inpatient department, emergency department, etc. More details regarding single unit simulations are available in the following articles: on a) emergency departments,[30-34] b) inpatient facilities[35-37] and, c) outpatient clinics.[38-40]

There has been a recent increase in the work on modelling: a) multi-facility units focusing on a single patient type, b) whole facilities with the focus of analysis being a particular subsystem, and c) facility subsystems serving multiple patient types.[26] We discuss example studies of each type here. Of the first type, Rewankar and Ward[41] developed a DES model for patients suffering from acute bacterial skin and skin structure infections. The model traced the treatment pathway of each patient through different departments - emergency department (ED), inpatient department, and outpatient department. Hamid et al.[42] focus on patients requiring elective open-heart surgery, and develop a two-stage optimisation and simulation approach to first mathematically determine optimal surgery schedules for the operating room, and then using DES determine the minimum number of beds in the downstream ICU to ensure an adequate patient service level. Of the second type, Lowery[43] developed a simulation model of a tertiary care hospital with a focus on the surgical suite and critical care area. The objective of the work was to determine the optimum number of beds in the critical area of the community hospital. The simulation model was designed to represent the arrival of patients to, and their flows through, nine different units in the modelled hospital. Similarly, Grida and Zeid[44] developed a systems dynamics simulation model of a medium-sized hospital with the focus of their analysis being identification of throughput improvement policies for the surgical department via a theory of constraints approach. Of the third type, Hasan et al.[45] developed a DES of an ICU in a hospital catering to patients of multiple types from upstream hospital units such as the emergency department, elective surgery, and emergency surgery. The objective of the study was to find suitable admissions and discharge policies to improve both patient and provider outcomes.

There is also a growing body of literature associated with the simulation of multi-disciplinary clinics, which are healthcare units established to provide integrated care from multiple care disciplines to patients with a given condition.[46-48] This is similar to the work by Rewankar and Ward,[38] but is not limited to units within a larger facility (that is, they can be standalone facilities as well). Examples of such studies include a simulation model of an 'integrated practice unit' for treating patients with lower extremity pain,[46] and a modelling study of a multi-disciplinary clinic for treating basal cell carcinoma.[47]

With respect to the above studies, there appears to be limited literature regarding whole facility simulation models that cater to multiple patient types with distinct clinical and operational flows through the facility (similar to PHCs). This is likely because most healthcare DES studies are undertaken to help analyse and/or solve specific problems associated with a facility, whereas our study aims to contribute towards establishing the computational infrastructure required to analyse the public health system in a region. Hence a key research contribution of our work involves developing a more broad-based simulation of the entire set of medical care components within the facility which incorporates all major patient types and their operational patterns within the simulation.

The widespread application of DES in healthcare implies substantial scope for its use in modelling and improving primary healthcare systems as well. Example studies include the use of DES for design of appointment scheduling systems for outpatient clinics which see multiple types of outpatients,[49] and the investigation of interactions between appointment scheduling policies and capacity allocation policies in an outpatient clinic with two patient types.[50] However, as mentioned above, studies concerned with modelling and simulation of single primary care facilities handling multiple types of patients – in particular, a mix of inpatients and outpatients – appear to be scarce. We encountered only one article that reported the use of DES to assess the impact of upgrading primary health care centres into bigger family health units (FHUs).[51] The authors modelled four types of consultations viz. a) medical, b) emergency or acute, c) nursing type 1 which mainly included diabetes and child or maternal care, and d) nursing type 2 consultations dealing with vaccination and other types of nursing treatments. It is unclear as to whether the study considered inpatient care.

From the standpoint of the scope of services included in primary care facility simulations, our research contribution here involves the inclusion of limited inpatient care and childbirth care in our PHC models inaddition to modelling general OPD consultations and emergency cases, as in the study by Fialho and Oliviera.[51] Note that, as discussed in Section 1, limited inpatient care and childbirth care services are likely to be offered in such primary care facilities in developing nations given that access to more comprehensive and specialised care is likely to be limited in semi-urban and rural regions.[52]

Given our use of the generic modelling approach to develop the PHC simulation models, we also briefly discuss the related literature here. Many articles describe the development of generic/reconfigurable simulations in a general context,[53-58] and/or in healthcare settings.[27,57,59,60] The above articles discuss the development of generic/reconfigurable DES models for physician clinics,[59,60] generic hospital simulation models,[27] and that of their subunits.[58,61] However, our search of the literature did not yield a study that demonstrated a generic modelling and model reuse approach for primary care centres that served multiple types of outpatients and inpatients. Further, we did not identify a study that categorised the surveyed facilities into different operational configurations and demonstrated, after the generic model is developed, the adaptation/reuse of the generic model to generate simulation models of these configurations.

A key research contribution of this paper thus involves addressing the above gap in the literature by providing a detailed demonstration of the implementation of the generic modelling approach to develop the PHC simulation models. In addition, we also demonstrate the adaptation of the archetypal PHC model to reflect the government mandate for PHC operations and idealised patient demand and outpatient consultation durations, and compare the performance of the PHC configurations in actual operation to the performance of this benchmark configuration.

## 2.2. Indian Context

There is very limited literature available regarding modelling the delivery of public healthcare in India. Most existing articles report on infrastructure, cost of delivering healthcare services, shortages of medical personnel in primary and secondary care hospitals, customer satisfaction, and out of pocket expenditures. A review article by Pandve and Pandve,[1] on primary healthcare services, describes the evolution of primary healthcare in India since independence.

Prinja et al.[62] reported the total annual cost of delivering health services at the PHC and CHC level. Their research determined the per capita per year cost of the complete package of healthcare services delivered at a PHC and estimated it to be INR 170.8. The availability of infrastructure and personnel in the PHCs was studied in the work of Sriram[8] for a district in the state of Andhra Pradesh. The author randomly selected fifteen PHCs and compared the data with the standards mentioned in the Indian Public Health Standards guidelines.[10] The work revealed that PHCs are deficient in both the human resources and the infrastructure required for day to day operations.

Mital[63] conducted a queueing study for resource planning associated with medical staff and inpatient beds in a medium sized hospital. The author used multi-channel queueing models parameterised by patient arrival and service time data to compute average utilisation estimates for inpatient beds and average lengths of stay were for male, female, and maternity wards.

In the Indian context, our literature search did not yield any study that computationally examined: a) PHC operations, and b) how their operational performance would respond to changes in demand and/or capacity. Further, there appears to be very limited healthcare facility simulations in general in the Indian context. Our study aims to address these gaps.

## 3. Model Development

In this section, we describe the development of DES models of PHC operations via the generic modelling approach. The DES models simulate provision of care to the following patient types: a) outpatients, b) inpatients and/or emergency cases, c) childbirth patients, and d) antenatal care patients.

The resources in each PHC consist of doctors, nurses, the pharmacist, the laboratory with the laboratory technician, and inpatient and childbirth beds. Each resource is accessed by one or more of the above patient types. The number of doctors varies between one and two depending upon PHC configuration, as some PHCs have two doctors while others operate with a single doctor. Staff nurses are also categorised as resources. Further, the staff nurses are divided into two categories: a) noncommunicable disease (NCD) trained staff nurse, who is present only during the OPD hours for conducting point-of-care tests and counselling related to NCDs (especially lifestyle diseases such as type 2 diabetes, hypertension, etc.) for patients above the age of 30; and b) the staff nurse, who attends to inpatients or emergency cases, and assists childbirth cases. Additionally, the in-house medical laboratory and pharmacy, with associated personnel, also remain available only during the OPD hours. We now describe the generic modelling process for the development of the PHC simulation models, including the operational data collected from PHCs.

### 3.1. Generic Modelling Approach and PHC Operational Data Collection

The generic modelling approach is a natural choice for developing simulation models of the PHCs. This is because the diverse health landscape of India implies that developing a broadly representative model of PHC operations would require surveying multiple instances of the facility of interest, identifying operational commonalities (and differences), and then conceptualising and developing this archetypal model based on the information synthesised from the survey - essentially the generic modelling approach. We provide a brief overview of the generic modelling approach and the concept of model reuse below, and place our PHC modelling effort within this context.

In their paper regarding generic modelling in the healthcare facility simulation context, Fletcher and Worthington[57] propose a classification scheme for a simulation model based on the extent to which it effectively represents multiple facilities – that is, for determining the extent to which a simulation model is generic (referred to in the paper as a model's 'genericity'). The authors suggest that the evaluation of a simulation model in terms of 'genericity' must be done in terms of two key attributes: model abstraction and transportability, and model reuse. A simulation model may possess one of four levels of 'genericity' in terms of model abstraction and transportability. These are, in descending order of genericity: level 1 - generic principle models not specific to an industry or a particular setting, such as general queueing models; level 2 - generic modelling frameworks or toolkits with models of units common to a specific industry (e.g., inpatient wards at hospitals, operating theatres), which can be leveraged to generate models of facilities of multiple types; level 3 - a generic model of a specific facility or process type (such as a generic model of accident and emergency (A&E) departments in the UK public health system, or outpatient clinics); and level 4 - models of a specific facility or process in a specific setting. In our case, the PHC models we develop clearly are of level 3, which also is the most commonly seen type in the literature - for example, the generic A&E model developed by Fletcher et al.[64]

The notion of simulation model reuse was explored - albeit very briefly - in the context of generic modelling by Fletcher and Worthington,[57] and in detail by Robinson et al.[65] Robinson et al.[65] postulate a spectrum of model reuse, with the following key types: code segment reuse, function reuse, model component reuse, and full model reuse. They also note that reuse of a model may be done for similar purposes as the original instance (e.g., a generic PHC model developed for broad-based operational analysis may be adapted to model and analyse operations of a specific PHC encountered in a different location), or for a different purpose (e.g., to simulate implementation of patient referral or diversion mechanisms across a network of PHCs, as in Fatma and Ramamohan[66]). Note that model reuse - in particular, component and full model reuse - does not necessarily imply reuse completely devoid of modification. In fact, Robinson et al.[65] suggest comparing the cost of adapting a model for reuse against that of *de novo* model development prior to opting for reuse.

In this context, we have developed a level 3 generic model of PHC operations, intending reuse in the same setting as well as in different settings, as well as reuse for both similar and different purposes. This also ties in with Fletcher and Worthington's[57] division of level 3 generic models into levels 3A and 3B, depending on the purpose and desired level of use. Level 3A models are meant to provide overarching insights regarding the facility's operations, and are intended for use by central planning stakeholders; whereas level 3B models are developed with multiple uses in mind (e.g., they can be adapted to model local instances of the facility) and hence may possess a higher degree of transportability. Overlap between these two types is possible, perhaps even desired, and the generic model we have developed achieves this overlap. We demonstrate this in the following sections where we describe experiments using the generic model to identify operational improvements in high-demand conditions that can be implemented on an overarching basis to existing and new PHCs regardless of configuration, and at the same time also modify the generic model to represent the PHCs with different operational configurations that we encountered in our data collection process. In Section 5, we also briefly discuss another case of reuse in the same setting, but for a different purpose - a use case relevant to the COVID-19 pandemic. In addition, even though the genericity of the PHC model that we develop extends only to public primary health facilities in the Indian context, reuse in different settings is also possible.

As described in Section 1, because of similarities in hierarchical public health systems in the developing world, we anticipate that the generic PHC model can be considered for adaptation and reuse to model equivalent facilities in these settings as well.

Developing a generic model of a public health facility typically involves the following steps: a) surveying a set of instances of the facility under consideration, involving operational data collection (patient flows, number of resources of each type, interarrival and services times for resources, etc.); b) identifying operational commonalities across the facilities surveyed, and conceptualising the operational structure of the generic model based on these commonalities; and c) parameterising, programming, and validating the generic model. In the following paragraphs, we describe the implementation of each step in developing the PHC models, and begin with describing our data collection visits.

We visited nine PHCs (out of ten total) in a north Indian district with a mix of urban, semi-urban, and rural populations to collect data regarding PHC operations. Permission to visit these PHCs and collect operational data was obtained from the district civil surgeon. Data collected included operational patterns (e.g., patient flow), staffing and resource levels (number of doctors, nurses, inpatient beds, etc.), patient arrival rates, and service time rates for different resources and patient types (doctors, outpatients, inpatients, clinical laboratories, staff nurses, etc.). This information is presented in Tables 1 and 2. Table 1 provides staffing information, the number of subcentres associated with each PHC, and information regarding the services rendered at these facilities. Table 2 summarises the data collected regarding service times per patient for different resources (e.g., doctor, laboratory). We present this information here as we were unable to identify literature that provided this (operational) data in the Indian context, and hence we anticipate that this information will benefit other researchers working on PHC and/or public healthcare operational policy.

Despite possessing institutional endorsement for the operational data collection, and also obtaining permission to collect PHC operational data from the district civil surgeon, we faced certain challenges regarding data collection during our visits. In most PHCs, staff were not fully cooperative, and did not allow access to the PHC premises to the extent required to collect comprehensive datasets for patient loads, service times, etc. This discomfort was likely due to their unfamiliarity with research personnel seeking to observe precisely their service patterns, and was alleviated only to a small degree by assurances that the data is anonymised. Further, because the precise entry and exit times of patients arriving to these PHCs and their subunits (e.g., time spent in the childbirth bed) was not maintained by the PHC administration, we recorded service times for multiple resources with a stopwatch in PHCs where we were provided the requisite access. Data collection in this manner was not possible for inpatient and childbirth care, and hence these service times were estimated based on discussions with doctors and nurses. However, even for outpatient services, we were not allowed to record associated service times in some PHCs (PHCs 7, 8, and 9). Further, even in PHCs 1 - 6, we were not able to record more than 10-15 observations at each resource before the staff requested us to stop. We faced similar restrictions in collecting data regarding outpatient interarrival times as well. Therefore, for patient load data, we were either provided brief snapshots of handwritten records regarding daily outpatient loads at a few PHCs, or at PHCs where such access was not provided, the patient load data was determined based on discussions with the medical staff (doctors, nurses, and pharmacists). For example, at one of the PHCs, we were provided access to daily outpatient loads recorded for 5-6 days, and at another PHC, we were only told that the outpatient load varied between 120-150 patients per day. Due to this, we were unable to observe - and hence could not incorporate in our models - any seasonal variation in patient load, or thinning effects (decrease in patient load as operating hours near closing time) that might be present in the PHCs. However, given that we capture the basic operational flow in the PHCs and the overarching patient loads and service rates are captured in the input parameters, further refinements regarding seasonal or weekly variations in patient load, thinning effects, etc. can easily be incorporated.

Table 1. Data summary of staffing level, patient load, and other facilities at PHCs

| PHC Visited | No. of Doctors | No. of Nurses | Patients /day | Monthly childbirth load | 24X7 facility? | No. of associated SCs | Laboratory Technician | Pharmacy manager |
|---|---|---|---|---|---|---|---|---|
| PHC-1 | 2 | 4 | 80-100 | 20-40 | Yes | 7 | 1 | 1 |
| PHC-2 | 1 | 4 | 50-70 | N/A* | No | 5 | 0 | 0 |
| PHC-3 | 2 | 4 | 60-80 | N/A* | Yes | 6 | 1 | 1 |
| PHC-4 | 1 | 4 | 35-60 | 5-20 | Yes | 8 | 0 | 0 |
| PHC-5 | 2 | 6 | 120-140 | 15-25 | Yes | 7 | 1 | 0 |
| PHC-6 | 2 | 4 | 30-50 | 10-12 | Yes | 6 | 0 | 1 |
| PHC-7 | 2 | 4 | 50-80 | N/A* | No | 3 | 0 | 0 |
| PHC-8 | 1 | 3 | 60-80 | 15-20 | Yes | 7 | 0 | 1 |
| PHC-9 | 2 | 3 | 120-150 | 30-50 | Yes | 5 | 1 | 1 |

*N/A = not applicable. PHC does not handle childbirth patients. No. = number.

We note that in certain PHCs, key operational and/or medical staff were not available. For example, four PHCs did not have a pharmacist, and five functioned without laboratory technicians. The four PHCs which did not have a pharmacist, as shown in Table 1, operated the pharmacy with the help of the staff nurse or the auxiliary nurse-midwife associated with one of the subcentres associated with the PHC.

In Table 2, the consultation time for doctors – the time the doctor spends with outpatients - was recorded for most PHCs using a stopwatch. Overall, 60 observations made during the OPD hours were used for determining the distribution of the doctor's service time. Similarly, observations for the time spent by the patients at the clinical laboratory for point-of-care tests and at the pharmacy were also recorded. More details regarding the parameterisation of the simulation using this data are provided in Section 3.4.

Table 2. Service time data summary for key PHC resources

| PHC Visited | Doctor (Seconds) Mean (SD) | Laboratory (Seconds) Mean (SD) | Pharmacy (Seconds) Mean (SD) |
| --- | --- | --- | --- |
| PHC-1 | 60 (18.9) | 232.6 (62.8) | 127.1 (13.4) |
| PHC-2 | 62.1 (25.5) | N/A | 142.8 (53.4) |
| PHC-3 | 53.2 (11.5) | 187.9 (33.3) | 160.7 (56.2) |
| PHC-4 | 47.5 (10.37) | N/A | 128.3 (39.2) |
| PHC-5 | 47.8 (13.6) | 200.5 (52.8) | 94.66 (33.3) |
| PHC-6 | 50 (9.5) | N/A | 91.4 (23.2) |
| Overall | 53.4 (12.2) | 207 (52.9) | 124.6 (51.8) |

*N/A = not applicable; SD = standard deviation.*

We now describe the conceptualisation of the generic model and other configurations based on the data collected above.

### 3.2. PHC Model Configurations and Parameter Estimation

It is evident from Table 1 that a variety of PHC configurations are currently in operation, and hence a single simulation model would not be able to capture this operational diversity. However, it was also evident that while PHCs differ in terms of staffing levels (e.g., number of doctors), services offered (e.g., presence/absence of childbirth services), and patient load (e.g, outpatient and inpatient demand), approximately the same operational pattern, is followed for a given patient type and service. Therefore, we developed a model of patient care operations in an archetypal or generic PHC, and modified the archetypal model to generate simulation models of other PHC configurations encountered in our visits. The archetypal model was created based on our observations of a set of PHCs that most closely resembled the guidelines for PHC operations prescribed by the Indian government.[10] Overall, we created three PHC configurations to broadly capture essential characteristics of the types of PHCs we encountered in our visits: one archetypal configuration, and two other configurations created by modifying the archetypal model. These configurations were developed based on key characteristics that affect operational outcomes: the number of doctors in the PHC, and whether they offer childbirth facilities and antenatal care. In addition, we have also created a configuration corresponding more closely to government-mandated PHC operational guidelines. This is done to present a comparison of the operational performance of the PHC configurations encountered during our visits and a configuration conforming more closely to government-mandated guidelines (which the archetypal/generic model also does), but with demand more closely following publicly available disease burden data, and doctor consultation times for outpatients closer to those observed internationally and in private facilities. We henceforth refer to this configuration as the 'benchmark' configuration. Note that this benchmark configuration differs from the archetypal configuration only in the outpatient load and the doctor's consultation time for outpatients. However, we still consider it to represent a separate configuration because the process of estimating the patient load for this configuration was significantly different and more involved than for the other configurations. Further, because we consider this configuration to represent an idealised benchmark case, we also assume higher consultation times for outpatients with the doctors - as will be discussed in Section 5, larger service times have been found to correlate with perceptions of higher quality by patients. More details regarding the patient demand estimation process are provided in Section 3.2.1. The essential facts regarding these configurations are presented in Table 3.

Configuration 1 represents the archetypal PHC operational pattern, as it is closest to the government mandate, with two doctors, daily observed outpatient load similar to demand estimated based on national disease incidence data (for more details, see Section 3.2.1), and provision of childbirth and ANC facilities. Further, it represents the superset of services and resource levels associated with other PHC configurations, and hence we decided to designate this configuration as the generic/archetypal model. PHCs 1, 5, and 9 from Table 1 can be considered as being represented by this configuration. PHC 6 may also be represented by this configuration if resource levels alone are considered; however, the patient load at this PHC is unusually low. Configuration 2 is developed to represent the cases where only one doctor is operating with a relatively lower patient load. These faciltiies also provide care to childbirth and

ANC patients. PHCs 4 and 8 can be considered as represented by this configuration. Configuration 3 was developed to represent cases where childbirth and ANC care facilities are not present. Given the low patient load at these PHCs (PHCs 2, 3, and 7) and the fact that the PHC guidelines[10] prescribe that only a single doctor is required if childbirth case load is less than 20 patients a month, we assumed that only a single doctor would operate in this configuration, similar to the case of PHC 2. We now describe the development of the benchmark PHC configuration.

### 3.2.1. Benchmark PHC Configuration

A key difference between the configurations described in the previous sections and the benchmark configuration is the patient load. Given that the PHC was set up to handle primary care in India, and that only up to 30 percent of current demand is addressed in public facilities, it is reasonable to assume that primary care demand at a PHC would be substantially larger than the average observed demand under conditions of greater trust in the public healthcare delivery system. Hence we have modelled the benchmark configuration to be a Type-B PHC, i.e. with two doctors. Hence, for this configuration type, in addition to two doctors, we assume one NCD nurse, four staff nurses working individually in consecutive eight-hour shifts, one laboratory technician, one pharmacist, and six inpatient beds and one labour room available 24 hours a day. PHCs are expected to have a minimum attendance of 40 patients per day per doctor per the PHC guidelines,[10] but these guidelines do not provide any information regarding the actual demand placed on the PHCs.

Table 3. PHC configurations

| Configuration | OPD/IPD/Childbirth/ANC interarrival time (minutes) | Number of doctors | Number of nurses[1] |
|---|---|---|---|
| Configuration-1 (generic) | 4/2880/1440/1440 | 2 | 4 |
| Configuration-2 | 9/2880/2880/2880 | 1 | 4 |
| Configuration-3 | 9/2880/NA/NA | 1 | 4 |
| Configuration-4 (benchmark) | 3/2880/1440/1440 | 2 | 4 |

[1]*Note: the nurses work in shifts - that is, each nurse works alone in an eight-hour shift. NA = not applicable. All configurations have 6 inpatient beds and 1 childbirth room (with a single bed).*

Hence we estimated the demand for primary care using morbidity data from the Brookings India report.[67] We used the disease incidence data reported for a ten-month period in India from the Brookings report[67] to estimate the number of people seeking medical care in the district where the

PHCs we visited were located. Next, using the percentage contribution of each disease to the total morbidity and the diseases that can be addressed at PHCs (identified by consulting PHC doctors), we estimated the daily demand for primary care. However, given that patients can visit PHCs, CHCs and DHs for receiving primary care, we then needed to estimate the fraction of the primary care demand that was addressed at PHCs. In the absence of data from the literature for estimating this, we assumed that the primary care demand is equally distributed among PHCs, CHCs, and the DH. We made this assumption because even through CHCs and the DH provide secondary and tertiary care (respectively) in addition to primary care, they have greater capacity as well, in terms of both the number of medical personnel and physical infrastructure (e.g., beds, larger premises). The total demand for primary care for the entire district was estimated as 8,560 patients per day, distributed uniformly across all the facilities. This yielded a patient load of roughly 570 per day seeking primary care at each PHC. However, considering only approximately thirty percent of the population utilise public healthcare facilities,[68] the final estimated patient load is approximately 170 patients per day.

We estimated the annual childbirth load a PHC may experience based on the birth rate for the district under consideration. Further, we then used data (37.6% births delivered in public hospitals) from the National Family Health Survey[69] to estimate the number of deliveries in public hospitals. We then assumed that out of these public facilities PHCs get only 20 percent of the birth cases because: a) PHCs are mostly located in the rural/remote areas with sparse population, and b) CHCs and DHs are better equipped in terms of facilities and staff and are located in relatively more thickly populated areas. This assumption is in line with the observations made during our visits, wherein we witnessed relatively low childbirth load in the PHCs. The estimated childbirth load was approximately 1 childbirth case per day. Additionally, in the absence of information in the PHC guidelines regarding inpatient load, the inpatient interarrival time is taken to be one per two days because, in keeping with the modelling assumption of higher-than-observed demand for the benchmark configuration, we assumed it to be slightly greater than the observed load (8 – 12 per month).

With regard to the consultation time with PHC doctors in the benchmark configuration, there is substantial variation seen across the globe. Studies report that consultation time with primary care doctors varied from 43 seconds in Bangladesh to nearly 22 minutes in Sweden.[70,71] Irving et al.[70] claimed that in 18 countries, comprising half of the global population, mean consultation time with primary care physicians was less than 5 minutes per patient. Given that we develop this configuration to represent a benchmark in terms of quality of care as well, we set the consultation times with doctors to be higher than that actually observed, because of correlations of outpatient consultation durations with patient perceptions of quality of care at PHCs (more details in Section 5). However, in India, because high consultation durations (e.g., exceeding 10 minutes) are likely to be difficult to implement due to the high demand, we have assumed the consultation time to be normally distributed with a mean of 5 minutes and standard deviation of 1 minute with a lower bound fixed at 2 minutes. Finally, the duration of post-childbirth stay in hospital is adopted from the PHC guidelines[10] in which a minimum stay of 48 hours is required. However, during the discussion with the doctors and the nurses we found that childbirth patients after the delivery rarely stay for more than 24 hours in the hospital and in general their length of stay lasts between 4 hours to 24 hours. Consequently, for our model, we have used a uniform distribution between 4 hours to 24 hours of stay. The nurse, laboratory, and pharmacist service time distributions and the inpatient bed length of stay are assumed to be similar to that estimated from the data collected during our PHC visits.

### 3.3. Patient Flow

We now describe the patient flow in the archetypal PHC (configuration 1). Figure 1 shows the patient flow for the archetypal PHC. All PHC resources - doctors, NCD nurses, staff nurses, pharmacy, the laboratory - are shared by all patient types, where applicable.

### 3.3.1. Outpatient Department

All the outpatients first go to the OPD room for a consultation and then are directed to the laboratory or to the pharmacy accordingly. In the OPD room, patients who are thirty years of age or above are directed for NCD related checks with the NCD nurse before consulting the doctor. Patients of age less than thirty years consult with the doctor directly. NCD checks involve recording patient medical history - blood pressure, body temperature, and weight measurements - and on some occasions, providing diet counseling and other such instructions.

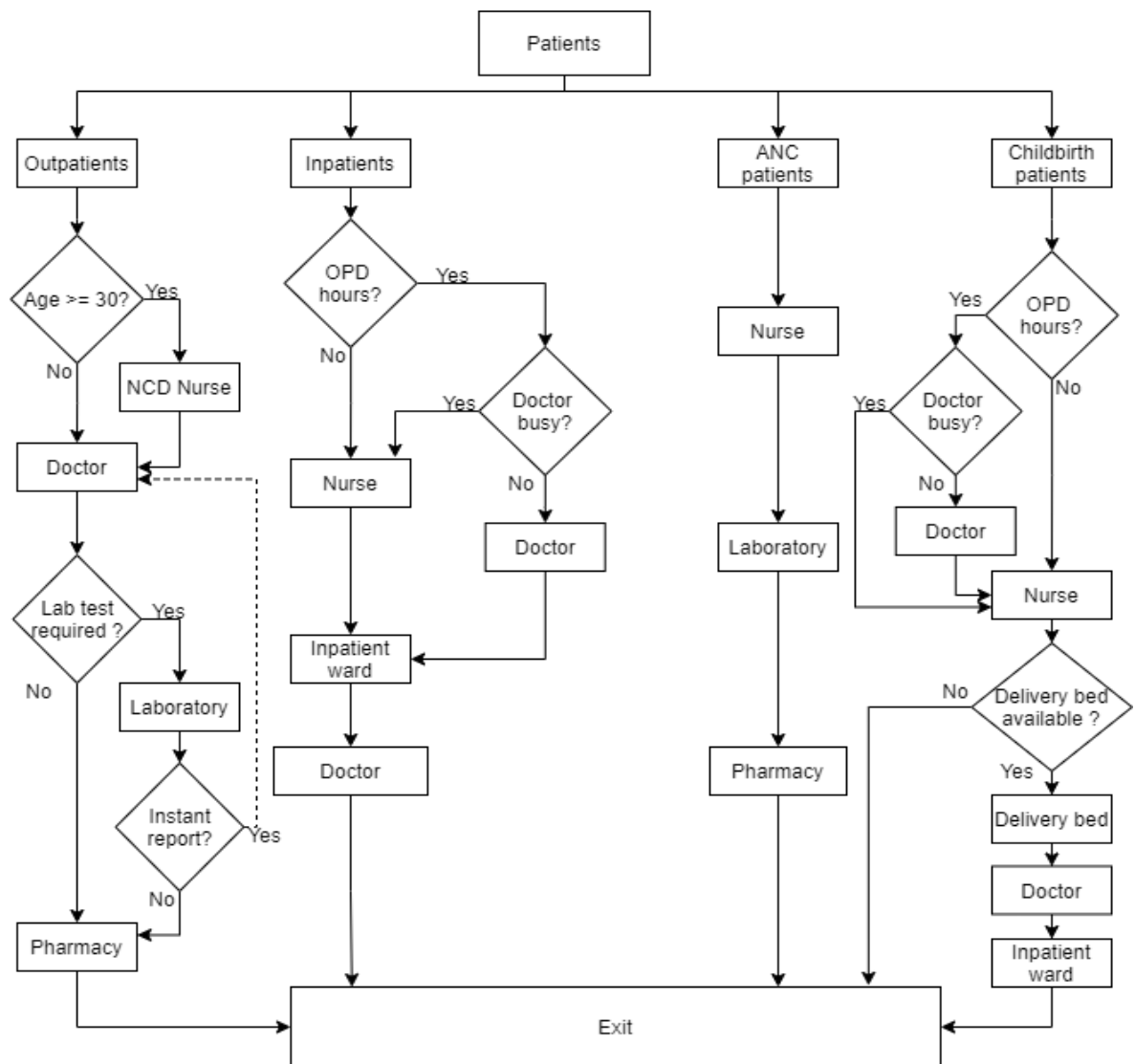

Figure 1. Patient flow in the archetypal PHC.
Note: the doctor and the nurse are shared between all patient types (where applicable), with childbirth patients and inpatients having nonpreemptive priority over outpatients.

After consulting the doctor, patients either go to the laboratory for tests, or they exit the PHC through the pharmacy. At the laboratory, two kinds of tests are typically performed: tests for which reports are generated in approximately 5-10 minutes, and others for which more than a few hours are required to generate reports. These are collected at a later date, and the associated patients are treated as new patients when they visit again since they are required to follow the regular flow in the facility. Patients requiring tests of the former category immediately go back to the OPD and leave through the pharmacy after registration as depicted by the dotted line in the figure and the latter group of patients return for consultation the next day. A point of note here is that those outpatients who require laboratory tests do not spend more than a few seconds when they first consult the doctor. This means that the doctor sends these patients immediately upon arrival to the laboratory for conduct of their tests, typically based on prior history with the patient. It is only when these patients collect their reports and return to the doctor (5-10 minutes after the test is conducted if it is a point-of-care test, or the next day if they undergo more complex tests) that the doctor conducts a full-fledged consultation. Therefore, even though patients consult the doctor twice during a single visit, the actual time spent with the doctor is effectively equal to that of a single visit.

Further, all outpatients invariably visit the pharmacy for registration (and provision of drugs if required) from where they exit the precinct. In the registration process, patient details are recorded, and a nominal fee (INR 5/10) is charged in some PHCs.

### 3.3.2. Inpatient Department

The PHC was established with the view to provide primary care and has no provisions for intensive inpatient care. Thus, the patients who are admitted on an inpatient basis comprise those with ailments that require care for less than 24 hours and if necessary are then referred to the CHCs or the DH. The average number of inpatient admissions varies widely across PHCs and also depends on the season – for example, the number of patients suffering from dengue fever or malaria increase during the monsoon months. When an inpatient arrives they first check whether the doctor is available. If the doctor is available, they are first attended to by the doctor and then by the staff nurse, and if not they are attended to by the staff nurse until the doctor becomes available. The length of stay for inpatients in PHCs is usually for a period of four to six hours and rarely exceeds eight hours. The nurse in charge of the IPD monitors the patient at regular intervals, provides medication, and maintains relevant records for each patient.

We note here that the inpatients have nonpreemptive priority in comparison to outpatients with regard to consulting with the doctor – that is, if the doctor is busy with an outpatient when an inpatient arrives, the inpatient moves to the head of the outpatient queue. The doctor then attends to the inpatient once they are finished consulting with the outpatient.

### 3.3.3. Childbirth Patients

According to the norms of the Ministry of Health and Family Welfare,[10] if the number of deliveries exceeds twenty per month, the PHC has to be provided with an additional doctor (i.e., total of two medical officers). However, based on our PHC visits, the implementation of this norm appears to be inconsistent, because some PHCs do not have childbirth facilities and still have two doctors, while others have only one physician. Further, despite the government mandate, it appears that not all PHCs provide childbirth services for all 24 hours, depending upon the availability of staff nurses.

When a childbirth patient arrives at a PHC, they check for the availability of a doctor. If a doctor is available, then the patient is attended to by the doctor and the staff nurse and is subsequently transferred to the labour room. Otherwise, the staff nurse attends to the patient and the doctor attends to the patient as soon as he/she becomes available. Once again, the childbirth patient has nonpreemptive priority for service from the doctor relative to the outpatient. However, childbirth patients and inpatients are served on a first-come first-serve basis by the doctor relative to each other. If the patient arrives during non-OPD hours, then the staff nurse attends to the patient and the patient is taken to the labour room. The staff nurse spends approximately 3 hours assisting the patient with the birth.

Since there is only one labour bed, the patient checks for the availability of the labour room. If it is unavailable for a period greater than 120 minutes then the patient is referred to another hospital. We have adopted 120 minutes as a threshold because in our case, it will not take more than two hours to reach the nearest public healthcare facility (PHC, CHC, or DH). Post-delivery the patients are shifted to the inpatient wards where they stay until they are discharged.

Note that inpatients and childbirth patient registration, dispensation of medication etc., is typically done during their length of stay in the inpatient ward or the childbirth bed, as applicable. Hence they do not exit the PHC via the pharmacy.

### 3.3.4. ANC Patients

Antenatal care is provided to pregnant women before childbirth. According to the Indian Public Health Standards (IPHS)[10] guidelines, pregnant women are advised to make four visits to the facility for routine examinations, medication, and counselling. The visit schedule is adopted from these guidelines wherein the first visit will be within twelve weeks of pregnancy, followed by the second visit between 14 and 26 weeks, third and fourth visits are scheduled between 28 and 34 weeks, and the fourth between 36 weeks and term. The staff nurse provides the antenatal care during these visits. In the simulation the next visit of an ANC patient is scheduled upon their arrival in the PHC for their current visit, and the number of visits they make is tracked to ensure it does not exceed four.

Upon the first visit, the staff nurse will make a registration card for the ANC patient and perform the predefined set of examinations and counsels the patient. ANC visits happen only during the OPD hours. Once ANC patients complete the examination, they undergo routine laboratory tests which, in most cases, are conducted in the PHC laboratory. They then exit the system through the pharmacy after collecting any medications or supplements.

## 3.4. Estimation of Simulation Parameters

In Table 4, we present model input parameter estimates with their associated probability distribution. We used 60 observations each for the doctor, pharmacy, and lab service duration, recorded using a stopwatch during our PHC visits, for estimating the associated input parameters. As described in Section 3.2, because we were unable to collect more than 10-15 observations of service time for a given resource, we had to pool service time observations across PHCs to obtain a sample of reasonable size. Goodness of fit tests for various distributions for each resource service duration were conducted on the Minitab statistical analysis software, and the Anderson-Darling statistic was utilised to identify the best-fitting distribution. The normal distribution best fit the service duration data for all three parameters (with negative values truncated at the minimum observed service durations during our visits; more details are provided in Appendix B.): doctor consultation time, pharmacy service time, and point-of-care tests at the laboratory. Note that the laboratory service time includes the time for interacting with the patient, collecting their sample, storing the sample, and recording patient and sample details. In other words, it represents the time between entry of the patient inside the laboratory to their exit. The laboratory reports for a given patient are generated approximately 5-10 minutes after sample collection; however, the time taken by the patient to pick up the report is neglible, and hence we do not include this within the laboratory service time. The service time distributions at the doctor, laboratory and pharmacy are provided in Appendix B.

The number of patients of each type (outpatients, inpatients etc.) arriving in different PHCs were estimated using the data maintained at the facilities, and also based on discussions with doctors and other associated staff. Patient arrival (for all four patient types, and for each configuration) is modeled by using an exponential distribution for the interarrival time. The average interarrival times (and consequently the average number of patients) at each configuration were estimated in the following manner. The number of outpatients visiting configuration 1 PHCs (PHCs 1, 5, and 9) range from 80 to 150 patients per day. These include patients visiting for the first time for a given case of illness as well as patients visiting for follow-up consultations on a previous case. Thus we assumed that approximately 125 patients visit on a given day for these PHC configurations, which include 90 first-visit patients, 20% patients on their first follow-up, and 10% visiting for their second follow-up, yield approximately 126 patients. Therefore, the interarrival time of 4 minutes at configuration 1 PHCs corresponds to first-time visits, with follow-up visits scheduled at the same time on any day between the next 3 and 8 days. With regard to the childbirth patient load at configuration 1, the number of cases range between 15 to 50 per month, and therefore we assumed the childbirth patient load to be 30 per month (close to the average of the range), corresponding to approximately one case per day. For estimating the inpatient load, we could access inpatient data load from only 3 PHCs, and in these, the average monthly patient load varied from 2 - 21 patients across an eight-month time horizon. Additionally, based on discussions with nurses and doctors across all PHCs, we determined that almost all PHCs experience low inpatient loads, ranging from 10 to 15 patients per month. Thus, we assumed that on average 15 patients will seek inpatient care at the PHC across all the configurations.

We modelled configurations 2 and 3 to have similar patient loads and resource levels (except for childbirth and ANC services) to illustrate the difference that offering childbirth and ANC services makes to operational outcomes. Hence, we discuss their parameterisation together. With regard to outpatient load at configurations 2 and 3, we see that their daily outpatient loads vary between 35 to 80. The outpatient load was therefore estimated to approximately equal to the midpoint of this range (approximately 55 patients per day, including follow-up visits), and hence the interarrival rate was also estimated in a manner similar to that configuration 1. The childbirth load at PHCs 4 and 8 (configuration 2) ranged from 5 - 20, and hence the childbirth load for this configuration was estimated to be close to the mean on the higher side, to be one case every alternate day.

Next, for assigning ages to outpatients (to determine which patients are required to visit the NCD nurse), we utilised Census 2011[5] data to estimate the proportion of the population aged less than 30 years. Thus, those aged 30 years and above were directed to the NCD nurse before consulting the doctor. Further, we assumed that an outpatient makes a maximum of two follow-up visits to the PHC after the first visit, considering the facility only provides primary care, in other words a patient can make a maximum of three visits to a PHC. Additionally, in the absence of published information regarding the proportion of patients requiring follow-up visits, we assumed that twenty percent and ten percent of the incoming outpatients would make two and three visits, respectively.

As for the inpatient, ANC, and childbirth cases, length of stay estimates were obtained from discussions with doctors, nurses, and also from relevant published data. The length of stay of inpatients is estimated based on discussions with the nurses and doctors because we could not access inpatient records for length of stay. For childbirth patients, considerable variation in the length of stay across facilities was reported by the nurses and the doctors. For these patients as well, because we could not access records for length of stay, we assumed a uniform distribution for the patient stay based on our interaction with the concerned staff in the PHCs. Also, the duration of labor for childbirth cases varies substantially from case to case.[72] In the data collection exercise, the doctors and nurses reported that the duration of labour could vary between 6-10 hours (assumed to follow a uniform distribution in the model) which we found to be consistent with the findings published in the literature.[73,74]

We did not encounter any ANC case during our visits, and the time estimates informed by the staff nurses were inconsistent and varied considerably from PHC to PHC. Hence, we have used estimates of ANC durations from the work of Both et al.[75] They measured the time taken per patient by the nurses for providing ANC services in their article. Similarly, for NCD checks we were able to record a very small number of observations because of staff apprehension that doing so would disrupt provision of care, and hence, we held discussions with the nurses to estimate their service duration.

Table 4. Facility independent input parameters

| Parameter | Value (Minutes) | Probability Distribution | Method |
| --- | --- | --- | --- |
| Doctor (OPD) consultation time | Mean = 0.87; S.D. = .21 | Normal | Data collection (Stopwatch) |
| Pharmacy service time | Mean = 2.08; S.D.= 0.72 | Normal | Data collection (Stopwatch) |
| Laboratory service time | Mean = 3.45; S.D.= 0.82 | Normal | Data collection (Stopwatch) |
| Nurse (NCD check) service duration | Min = 2; Max = 5 | Uniform | Data collection (Nurse discussion + limited observations collected) |
| Doctor (Inpatient) service time | Min = 10; Max = 30 | Uniform | Data collection (Doctor discussion) |
| Nurse (Inpatient) service time | Min = 30; Max = 60 | Uniform | Data collection (Nurse discussion) |
| Nurse (Childbirth) service duration | Min = 120; Max = 240 | Uniform | Data collection (Nurse discussion) |
| Inpatient bed length of stay | Low = 60, High = 360, Mode = 180 | Triangular | Data collection (Doctor and nurse discussion) |
| Labour bed length of stay | Min = 300, Max = 600 | Uniform | Data collection (Doctor and nurse discussion) |

| Doctor (Childbirth) service time | Min = 30; Max = 60 | Uniform | Data collection (Doctor and nurse discussion) |
|---|---|---|---|
| Childbirth patient bed length of stay | Min = 240; Max = 1440 | Uniform | Doctor and nurse discussion; IPHS guidelines[10] |
| ANC visits | Four visits | Deterministic | IPHS guidelines[10] |

### 3.4.1. Model Assumptions

Here we list the assumptions made for the simulation model.

- The outpatient unit runs for 6 hours per day and all the outpatients who arrive from morning 8 AM till 2 PM consult with the doctor.
- All the resources and staff remain fully available during operations.
- The performance of the staff, in terms of their service time parameters, does not change with time during shifts, and they are available throughout the shifts without breaks.
- There is one staff nurse per shift (eight-hour shift), thus in a day, three nurses work in tandem while the fourth nurse has a night off.
- Each nurse does administrative work of approximately one hour per shift.
- Pharmacy and laboratory are available continuously during the outpatient unit hours.
- Doctors do not attend to the patients (inpatient/childbirth cases) after outpatient unit hours.
- Doctors also perform administrative work - e.g., paperwork associated with running the PHC. Based on discussions with the doctors, the administrative work is taken as normally distributed variable with a mean value of 100 minutes and a standard deviation of 20 minutes.
- All the outpatients go to the pharmacy after consulting the doctor.
- We only consider patient care provided on a direct basis in our models. For example, doctors are responsible for organising various health camps and conduct field visits as part of public health outreach programmes, and because the nature of these programmes changes from period to period based on government policies, we do not include these in our studies.

## 4. Simulation Experiments and Results

The PHC simulation was programmed using the Python programming language on the Pycharm IntelliJ integrated development environment. Python's Salabim package,[76] which is a third-party package developed for discrete event simulation, was used in programming the model. The simulation was run for 365 days, with a warm-up period of 180 days. The warm-up period, per usual simulation practice, was run with the same set of patient arrival and service rates as in the steady state period (see Table 3 and Table 4). The duration of the warm-up period was chosen to allow a sufficient amount of time for the simulation outcomes to achieve steady state. Results from 100 replication runs were collected for all computational experiments. All computational experiments were performed on a workstation with a quad-core Intel Xeon processor, base frequency of 3.7 GHz, and 16 gigabytes of RAM. Completing 100 replications required approximately forty-three minutes and forty-four seconds. The software for the generic PHC model (configuration 1) is available at this location: https://github.com/shoaibiocl/PHC-/blob/main/PHC.py.

We begin by discussing our efforts to validate the results of our models and extract analytical insights related to queueing systems that form part of the PHC models.

### 4.1. Model Validation and Queueing Analysis

We were unable to perform external validation of the simulation model by comparing its outcomes to, for example, operational outcomes published in the literature, because we were unable to identify any previously published data in the Indian context regarding PHC operational outcomes such as average outpatient waiting time, utilisation of doctors, staff nurses, etc. However, the outcomes generated by the model for all configurations were in good agreement with those observed during our visits to the PHCs. For example, the waiting times observed for outpatients visiting configuration 1 PHCs were negligible, and the utilisation of all resources, as observed in terms was also well below 50%.

In addition, we were able to compare the estimates of time spent waiting in the outpatient queue and doctor's utilisation generated by our model to the corresponding estimate obtained from primary data collected from a visit to the primary care unit of a similar public health facility in another district. We measured the average time spent waiting in the outpatient queue before consulting the primary care doctor for 40 patients and compared it to the estimates generated by our simulation model for configuration 1, the PHC closest in operational patterns to the primary care unit facility. The observed waiting time was estimated to be approximately 20.03 seconds with a standard deviation of nearly 20 seconds, caused by the presence of a large number of observed waiting times of 0 seconds (42.5% of outpatients observed during our visit had 0 second waiting times, and the maximum waiting time observed was 84 seconds). The simulated average waiting times generated for configuration 1 was negligible (< 5 seconds), thereby indicating that our simulations appear to capture PHC operational outcomes reasonably well. Further, the doctor's utilisation was estimated to be approximately 20% during our visit, in comparison to the corresponding estimate of approximately 25% for configuration 1.

From the perspective of internal validation (i.e., ensuring the simulation was implemented correctly by comparing simulation outcomes to analytical estimates), we also compared the doctor's average utilisation estimates from the simulation models to the corresponding analytical estimates obtained using queueing theory concepts. In the subsequent analysis, we consider the utilisation of the doctor to be a random variable, to reflect the fact that in steady state operations of the PHC, the measurement of utilisation over different time periods will yield slightly different estimates of the doctor's utilisation. In steady state, we can assume that these estimates are $iid$, and have expected value $\rho_d$ and standard deviation $\sigma_d$. Given this view of the doctor's utilisation, we assume that the best estimate of $\rho_d$ can be generated by an accurate simulation of PHC operations run for an sufficiently long duration.

We computed the average utilisation of doctors, ignoring their administrative work, from the simulation models and compared the results with analytically computed average utilisation estimates. The parameters of the analytical queueing system from which we estimate the server's (the doctor's) average utilisation remain the same as that of the simulation model - i.e., the doctor serves outpatients, inpatients and childbirth cases, each with exponentially distributed interarrival times and corresponding general (non-exponential) service time distributions, as given in Table 4. The analytical estimate of the average utilisation of a server in such a queueing system is computed by summing the average utilisations estimated assuming each patient type was the only patient type arriving in the system.[77] This is given below:

$$\rho_a = \rho_o + \rho_i + \rho_c \qquad (1)$$

Here $\rho_a$ represents the analytical estimate of the average utilisation of the doctor under demand from three types of patients, and $\rho_o, \rho_i, \rho_c$ represent the average utilisation values for the doctor assuming the doctor handles demand for care from only outpatients, inpatients and childbirth cases, respectively. $\rho_o, \rho_i$ and $\rho_c$ are estimated in the usual manner; that is, as the ratios of the average service durations $\mu_o, \mu_i$ and $\mu_c$, respectively, and the corresponding average interarrival times $\lambda_o, \lambda_i$ and $\lambda_c$. The comparisons between the average utilisation estimates for the doctors from each configuration (denoted by $\hat{\rho}_d$) with $\hat{\rho}_d$ are summarised in Table 5 below. Note that the $\hat{\rho}_d$ were generated by assuming no outpatient revisits and administrative work to simplify the exercise. We conducted one-sample $t$ tests to check whether the analytical estimates of $\rho_d$ lay within the interval $(\hat{\rho}_d - k_\alpha \hat{s}_d, \hat{\rho}_d + k_\alpha \hat{s}_d)$, where $k_\alpha$ is chosen to reflect the maximum allowable deviation from $\hat{\rho}_d$. Thus we do not check whether the analytical estimates lie within a confidence interval associated with the sampling distribution of $\hat{\rho}_d$, and instead check whether it lies within an acceptable range around $\hat{\rho}_d$ within the distribution of the utilisation. We adopt this approach because in steady state, the value of $\hat{s}_d$ is very small (as expected), and hence checking against the sampling distribution of $\hat{\rho}_d$ would be unduly restrictive. The results of this exercise are summarised in Table 5.

We see from the results in Table 5, that the $\hat{\rho}_d$ and $\rho_a$ estimates are statistically similar for all configurations except for the benchmark case. Even in this case, the difference between the analytical and simulation estimates is < 4.0%.

Table 5 Internal validation outcomes for doctor's utilisation

| PHC Configuration | $\hat{\rho}_d$ | $\rho_a$ (p-value, % difference from $\hat{\rho}_d$) |
|---|---|---|
| Configuration 1 | 0.122 | 0.1155 (0.13, 5.6) |
| Configuration 2 | 0.109 | 0.1042 (0.26, 4.6) |
| Configuration 3 | 0.099 | 0.0991 (0.82, 1.00) |
| Benchmark configuration | 0.870 | 0.840 (0.02, 3.6) |

The above exercise for performing internal validation of our simulation outcomes also motivated us to develop two analytical approximations for the utilisation of the server in the queueing system represented by the doctor providing service to outpatients, inpatients and childbirth patients. The development, simulation-based validation, and avenues of use of these analytical approximations are described in detail in Appendix C.

While the analytical results in Appendix C. are applicable to general multi-class queueing systems where there are significant disparities between the arrival and/or service rates of one job class relative to others, they were developed based on our observation of the simulation outcomes for the queueing system represented by the doctor's service of outpatients, inpatients and childbirth patients. Therefore, the numerical verification of these analytical results using our PHC simulation can also be considered to be another level of internal validation of our simulation models - that is, if the simulations lacked internal validity, the verification exercise would have yielded results contradictory to those in Theorems C.1 and C.2.

## 4.2. Simulation Outcomes

The results from the simulation models of the as-is and benchmark configurations are presented in Table 6.

Table 6. Operational outcomes for each PHC configuration simulation

| Simulation Outcome | Configuration-1 (2/130/0.5/1/1/6/1)* | Configuration-2 (1/60/0.5/0.5/0.5/6/1)* | Configuration-3 (1/60/0.5/0/0/6/0)* | Benchmark Case (2/170/0.5/1/1/6/1)* |
|---|---|---|---|---|
| | Mean (SD) | | | |
| Doctor utilisation | 0.268 (0.003) | 0.372 (0.004) | 0.354 (0.002) | 1.142 (0.006) |
| NCD Nurse utilisation | 0.865 (0.011) | 0.469 (0.005) | 0.468 (0.005) | 1.232 (0.019) |
| Staff nurse utilisation | 0.323 (0.008) | 0.243 (0.006) | 0.16 (0.001) | 0.322 (0.008) |
| Pharmacist utilisation | 0.643 (0.004) | 0.288 (0.003) | 0.289 (0.003) | 0.855 (0.005) |
| Lab utilisation | 0.559 (0.008) | 0.254 (0.004) | 0.239 (0.004) | 0.736 (0.011) |
| Inpatient bed utilisation | 0.093 (0.004) | 0.055 (0.003) | 0.011 (0.001) | 0.093 (0.004) |
| Labour bed utilisation | 0.283 (0.01) | 0.153 (0.009) | Not applicable | 0.281 (0.012) |
| Mean length of OPD queue (number of patients) | 0 (0) | 0.007 (0.001) | 0.001 (0) | 0.817 (0.027) |
| OPD queue waiting time (minutes) | 0.009 (0.004) | 0.171 (0.032) | 0.034 (0.001) | 6.789 (0.268) |
| Mean length of pharmacy queue (number of patients) | 0.09 (0.002) | 0.01 (0.001) | 0.009 (0) | 0.15 (0.002) |
| Pharmacy queue waiting time (minutes) | 1.025 (0.021) | 0.244 (0.008) | 0.232 (0.006) | 1.282 (0.018) |
| Mean length of Lab queue (number of patients) | 0.094 (0.003) | 0.012 (0.001) | 0.011 (0) | 0.188 (0.001) |
| Lab queue waiting time (minutes) | 2.084 (0.054) | 0.606 (0.023) | 0.571 (0.02) | 3.135 (0.005) |
| Fraction of childbirth cases referred | 0.156 (0.019) | 0.088 (0.022) | Not applicable | 0.157 (0.18) |

*Number of doctors/OPD cases/IPD cases/childbirth/ANC (patiets)/inpatient beds/labour room

It is evident that all as-is configurations are substantially underutilised when compared to the benchmark configuration. In the case of the benchmark configuration, the higher demand (nearly

30% higher than configuration 1, the as-is configuration with the highest demand) and the higher average doctor service time for outpatients (more than 5 times that in the as-is configurations) is the main cause for the increase in resource utilisation. In the case of configurations 2 and 3, the increase in the doctor's utilisation despite the decrease in outpatient load is explained by the fact that only one doctor is present during the OPD hours. The operational implication of utilisation estimates exceeding 1.0 is that the resource under consideration spends time over and above their designated work hours in completing care provision to patients who arrive during their work hours. For example, the utilisation estimate of 1.142 for the doctor in the benchmark configuration implies that the doctor spends approximately 14% more time than their designated work hours in providing care to all patients who arrive during their work hours.

It is also clear from Table 6 that inpatient as well as labour bed utilisation values are low in all cases. Despite the low values of the labour bed utilisation, we see that a significant fraction of childbirth patients are referred elsewhere. This occurs because these patients happen to visit the facility while the labour bed is occupied by another childbirth patient and are referred elsewhere when their waiting time exceeds two hours. This could thus indicate that at least one of the inpatient beds could be converted into a second labour bed. We explore this in simulation experiments in the following sections.

The higher values of resource utilisations for the benchmark configuration are a cause for concern, as the patient demand was estimated assuming that only 30% of the total patient load is addressed in public facilities. If the proportion of patients seeking care at public facilities increases (for example, to 50%), then it is clear that the current capacity of the PHC (at an "ideal" mean service time of 5 minutes for the doctor) is not sufficient to effectively address the demand. This indicates a need for expanding either the capacity of the PHC in terms of adding sufficient medical personnel, exploring alternative operational patterns, or establishing new PHCs. We explore the first and second options in the following sections, as the third is outside the scope of the paper.

### 4.3. Sensitivity Analysis and Configuration Optimisation

In this section, we conduct sensitivity analyses for the generic PHC model – that is, the configuration 1 PHCs. The sensitivity analysis involves determining how PHC operational outcomes (e.g., average outpatient waiting time, resource utilisation) respond to changes in demand, which are modeled by varying outpatient, childbirth, and inpatient case arrival rates. The doctor's average service time for outpatients is varied from the default estimate of slightly less than one minute to 2.5 minutes and 5 minutes. We did not vary the outpatient service time beyond 5 minutes because, as discussed earlier in Section 3.2.1 service times comparable to that seen in developed nations (10 - 20 minutes per consultation) is unlikely to be viable at current capacity levels in the Indian context due to the high patient demand. Hence we increase the doctor's average outpatient consultation time to a maximum equal to that assumed for the benchmark configuration (5 minutes). We also consider an intermediate average outpatient consultation time of 2.5 minutes. Similarly, the outpatient arrival rate was varied to a maximum of 170 patients per day, equal to that in the benchmark configuration. The estimation of the outpatient arrival rate for this configuration is described in detail in Section 3.3.1. Similarly, for the second set of sensitivity analyses, the inpatient, childbirth and ANC patient arrival rates were at maximum doubled given that a significant proportion (approximately 16%) of childbirth patients were being referred elsewhere even at the current childbirth patient arrival rates. The results of these experiments are presented in Figure 2 and Figure 3. In these figures, we only present outcomes that change significantly when the parameters that are the subject of the sensitivity analysis are varied.

Figure 2. Sensitivity analyses for the configuration 1 PHC around average outpatient load at different doctor's outpatient consultation time (minutes). Note: S.D. represents standard deviation.

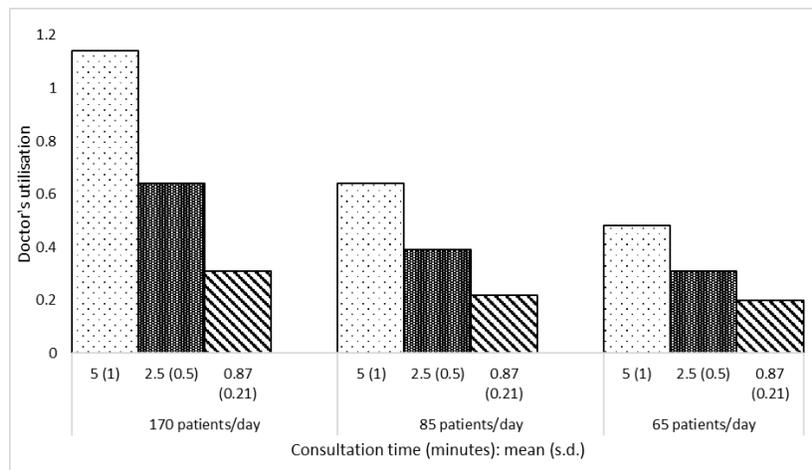

Figure 2a. Impact on doctor's utilisation.

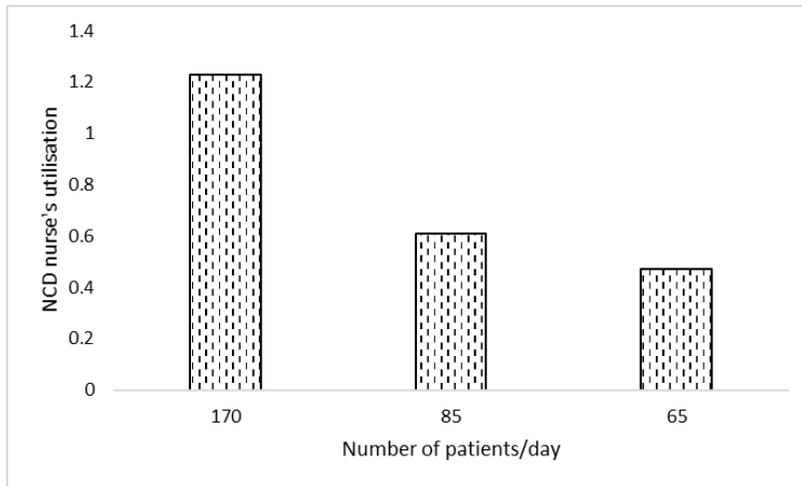

Figure 2b. Impact on the NCD nurse's utilisation.

*Note: the doctor's consultation time does not impact this outcome.*

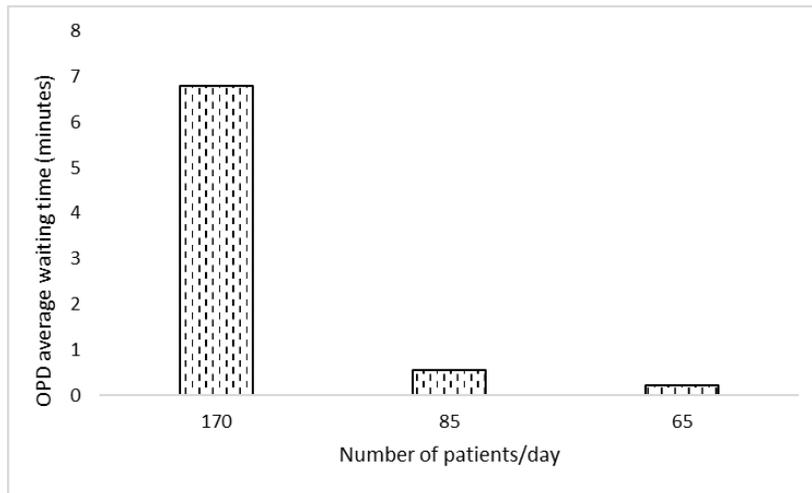

Figure 2c. Impact on the average waiting time (minutes) in the OPD.

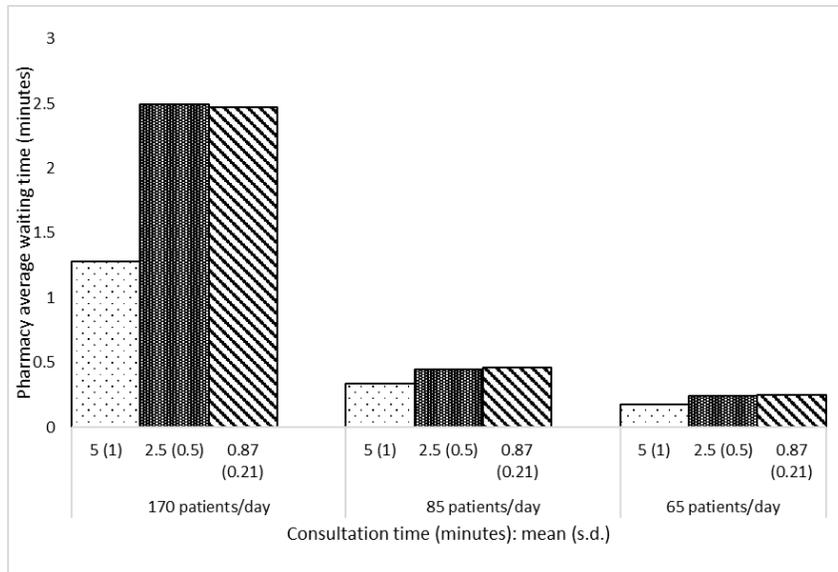

Figure 2d. Impact on the average waiting time (minutes) in the pharmacy.

Figure 3. Sensitivity analysis for configuration 1 PHCs around changes in inpatient/childbirth-patient/ANC-patient loads. Note: '*' represents the average number of inpatients/childbirth-patients/ANC-patients per day; S.D. represents standard deviation..

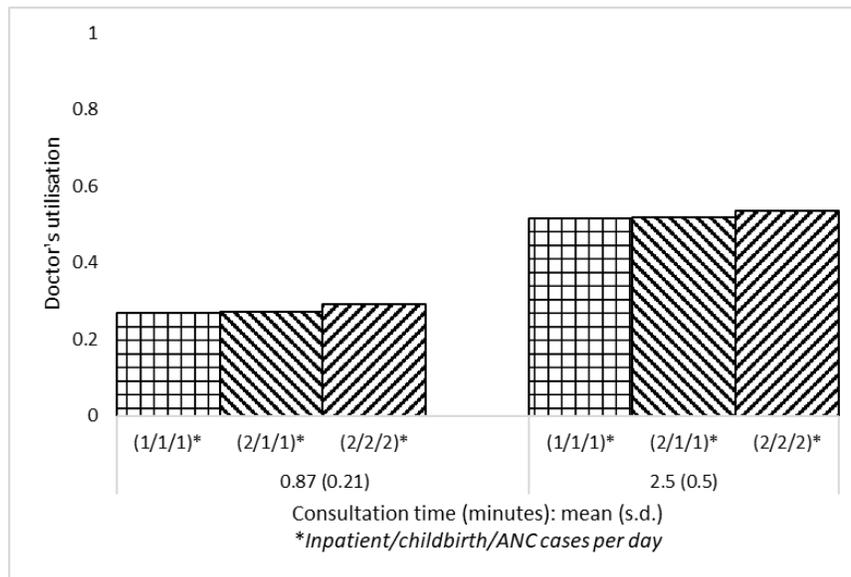

Figure 3a. Impact on the doctor's utilisation. Two levels of outpatient consultation times (minutes) are used.

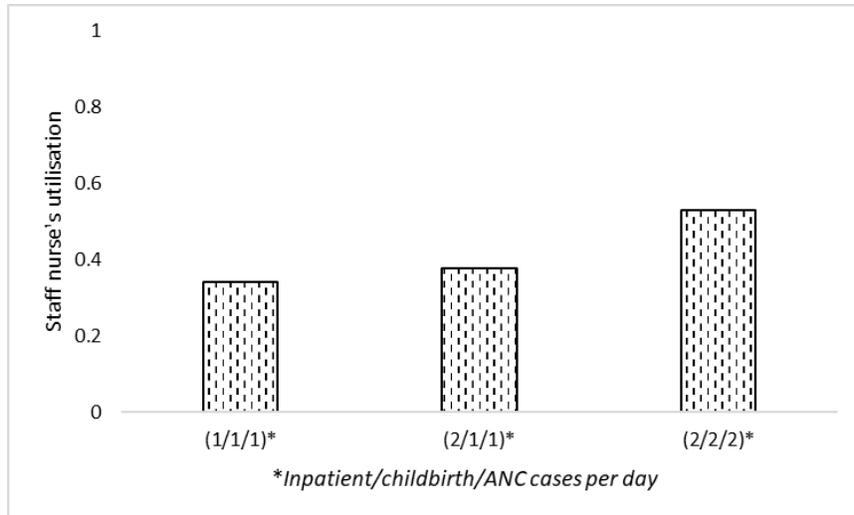

Figure 3b. Impact on the staff nurse's utilisation.

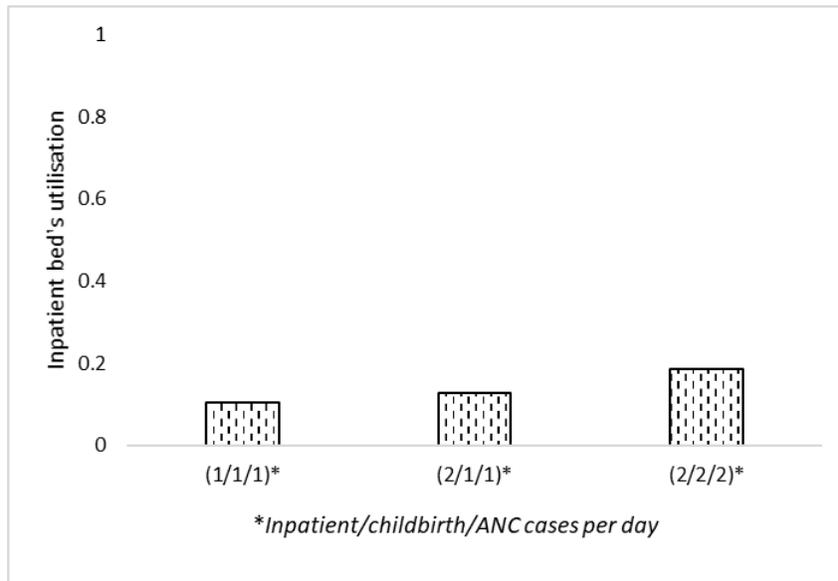

Figure 3c. Impact on the inpatient bed's utilisation.

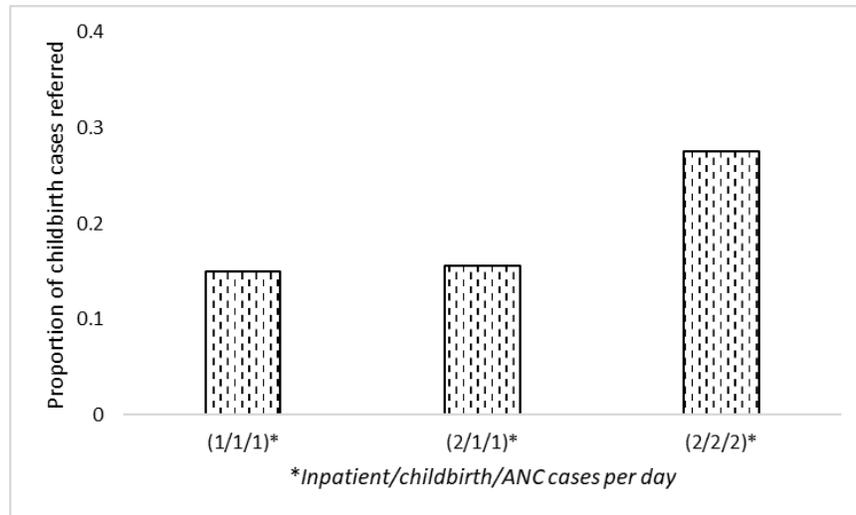

Figure 3d. Impact on the proportion of childbirth cases referred elsewhere.

As expected, resource utilisation increases with increases in outpatient load. However, we see that doctor and nurse utilisations (Figures 2a and 2b) exceed 100% only in one case – that is, when outpatient load is 170 patients/day and doctor service time is at its highest value (mean of 5 minutes per patient). This indicates that if maintaining a service time of 5 minutes is not feasible in the Indian context, increasing the average service time to at least 2.5 minutes from the current estimate (< 1 minute/patient) is well within current capacity limits. However, if it is imperative to maintain a 5 minute/patient mean service time, then more resources must be added to reduce doctor and NCD nurse utilisation. NCD nurse utilisation in particular is a cause for concern; however, a potential solution could involve having the staff nurse assist with NCD checks given their relatively lower utilisation. We explore this in simulation experiments described in the subsequent paragraphs.

Interestingly, at each outpatient arrival rate, as the doctor consultation time decreases the waiting time before pharmacy increases as indicated in Figure 2d. This occurs because at lower consultation times, more patients reach the pharmacy in a given time duration and thus results in slightly longer waiting times. However, since the number of patients remain the same, the pharmacist utilisation does not change.

From the sensitivity analyses depicted in Figures 3a – 3d, we see that increase in inpatient, childbirth and ANC case demand does not affect the doctor's utilisation significantly because the major portion of the doctor's time is consumed by outpatient demand. However, staff nurse utilisation, depicted in Figure 3b, increases substantially as the staff nurse is primarily responsible for attending to childbirth and inpatient cases. We also note that waiting times for outpatient-related resources (laboratory, OPD consultation, etc. - not depicted in Figures 3a – 3d) increase marginally because the associated resources are also required by inpatient/childbirth/ANC cases, which increase in number in the above scenarios. Also, as mentioned previously, the substantial proportion of childbirth cases that are referred elsewhere due to labour bed unavailability is cause for concern (27% when the average number of childbirth cases per day is increased to 2).

In the subsequent sections, we discuss some potential solutions through which operational outcomes for both medical personnel and patients can be improved.

### 4.3.1. Doctor's Utilisation

The sensitivity analyses reveal that at an average outpatient load of 170 patients per day, the utilisation of doctors increases substantially with increases in the mean service time and exceeds 100% at an average consultation time of five minutes/patient, implying that doctors may stay longer than the designated working hours to serve all patients arriving within the designated working hours. To address this, we experimented with letting the staff nurse (whose utilisation is approximately 32%) take over the administrative work. This led to a 12% drop in the utilisation level, which implied that the doctor's utilisation still exceeded 100%. Implementing this measure resulted in increasing the staff nurse utilisation to nearly 40%. Therefore, we then considered a situation wherein the staff nurses require minimal intervention in childbirth cases. We assumed that in 50% of childbirth cases, staff nurses require no intervention by the doctor; require only one-third of the typical amount of intervention in 30% of cases, and require full intervention in the remaining 20% of cases. This led to a decrease of the doctor's utilisation to 101% (a further decrease of approximately 1%), and an increase in the nurse's utilisation to 40%. The nurse's utilisation does not change significantly because the nurse attends to the patient irrespective of the presence of the doctor. Thus, while this reduction in administrative and childbirth work improves the doctor's utilisation, it does not address the issue entirely, as the doctor's utilisation remains at 100% (in stochastic conditions, utilisations of less than 100% are recommended).

Finally, we also investigated the effect of stationing an additional doctor in the PHC. This yielded an average utilisation of well below 100% for each doctor. This indicates that additional doctors can be rotated in from less busy PHCs (perhaps from configuration 2 PHCs) to a busier PHC when required.

### 4.3.2. Labour and Inpatient Bed Utilisation

In the simulation model, beds are divided into two categories: labour and inpatient beds. For inpatient ward beds, the utilisation is 7% to 10%. These are utilised by inpatients and the childbirth patients after the labour period is completed. However, as seen from sensitivity outcomes, changing inpatient and childbirth case arrival rates do not significantly increase their utilisation levels, with a maximum utilisation of 20%. We also observe that if the number of beds is reduced to four from six, the utilisation level is observed to be approximately thirty-three percent even under higher demand conditions (two inpatient and childbirth cases/day).

Labour bed utilisation is nearly 28% for configuration 1 PHCs. However, because there is only a single labour bed and labour bed utilisation times are relatively high, a significant fraction of patients are referred elsewhere. For minimising the number of childbirth cases referred elsewhere due to the occupied delivery bed, one of the inpatient beds was converted into an additional labour bed. The results of this investigation are depicted in Figure 4, which show the change in the fraction of cases referred elsewhere and labour bed utilisation by adding one extra labour bed at different childbirth patient loads. The results show a significant drop in the fraction of cases referred elsewhere; however, as expected, it is evident that as the childbirth case load increases, more labour beds will be required.

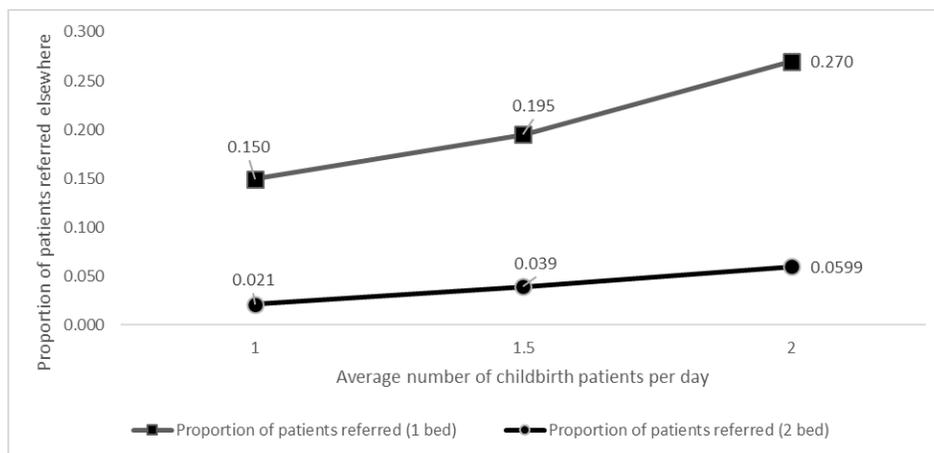

Figure 4. Effect of including additional labour beds on proportion of childbirth patients referred elsewhere.

### 4.3.3. NCD Nurse Utilisation

From the standpoint of nurse utilisation, NCD nurse utilisation presented the most cause for concern. We see from Figure 2b that NCD nurse utilisation exceeds 100% (123%) when outpatient interarrival time (iat) is 3 minutes, and at lower outpatient loads, utilisation remains at 61% (iat = 6 minutes), and 47% (iat = 9 minutes). NCD nurse utilisation can be addressed by a) having the staff nurse assist with the administrative work performed by the NCD nurse, and/or b) the staff nurse assisting with NCD checks for outpatients. When the administrative work alone is assigned to the staff nurse the average utilisation of the NCD nurse decreases to 100%. Further, in addition to the administrative work when the staff nurse assisted for NCD checks (for 10% cases) the utilisation of NCD nurse dropped to 71%. Thus, it could be a viable solution to assign administrative work to the staff nurses and also take their assistance for NCD checks, wherever possible, in the case of high demand PHC configurations.

## 5. Discussion & Conclusions

In this work, we study the operations of primary health centres via the method of discrete-event simulation. We develop the PHC DESs by visiting nine PHCs in a north Indian district and collecting operational data from each PHC. Our key findings from the PHC visits include the following: a) while operational patterns around provision of patient care are similar across PHCs (enabling the development of a generic PHC model), a variety of operational configurations in terms of services offered and medical staffing levels, beyond the two configurations mandated by the Indian government, appear to operate; b) not all PHCs follow the minimum staffing requirements mandated by the government; and c) the inpatient, outpatient and childbirth case load at PHCs appear on average to be lesser than demand estimated from disease burden data; however, significant variation in demand is also observed between PHCs. We also note that our visits were limited only to a single district. While we conducted limited additional visits to primary healthcare facilities in other districts for collection of data to validate the model and observed similar patterns, generalisation of our observations to PHCs across the diverse and vast health landscape of India must be made with appropriate caution. The generic modelling approach that we have adopted can provide some relief here - given the diversity in the operational configurations of PHCs that we encountered in this district (e.g., compare PHCs 4, 5 and 6 in Table 1), and the fact that the government guidelines for PHC configurations are applicable on a national level, it is likely that one of the configuration models that we have developed will be similar to PHC configurations encountered in other districts across the country.

Our simulation outcomes indicate that the medical personnel, with the exception of the NCD nurse, are underutilised in the PHC. This is likely due to both low demand conditions, and the high service rate of the medical personnel – in particular that of the doctors (average service times < 1 minute for outpatients). The low service times we have observed in our PHC visits is consistent with service times that have been observed across India.[70] Thus, if only medical care and administrative duties are considered, low utilisation conditions are likely to be encountered in other Indian PHCs as well.

A few studies have been published that investigate factors that affect patient perceptions of quality of care at public and private hospitals,[78] and in the relatively recent past, factors that drive why patients bypass PHCs.[79,80] Narang[78] studied the impact of a variety of factors affecting patients' perceptions of quality of health services in both public and private hospitals in a large Indian city. Prominent among the factors that patients perceived as important was the time spent by doctors at these facilities during consultations. The most influential factors, however, were adequate clinical examination and the compassion and respect shown by care providers to patients, with adequate clinical examination potentially correlated with sufficient time provided to patients. On all these factors, private hospitals performed better than public hospitals, with public hospitals performing better only on accessibility. Similarly, in a qualitative study by Ramani et al.[80] where the authors interviewed both care providers (doctors) and patients who accessed PHCs, patients expressed that insufficient attention was a key factor in their bypassing PHCs. Rao and Sheffel[79] investigated the impact of clinician competence and structural quality of PHCs (availability of drugs, physician absenteeism, etc.) on patients bypassing PHCs, and determined that increasing provider competence (measured by accuracy of diagnosis) and structural quality reduced bypassing only up to a certain extent. This was despite the fact that patient costs were half of that in private facilities. This suggests that in addition to accuracy of diagnosis, as determined by Narang[78] and Ramani et al.[80], sufficient attention and the manner of providing care may improve patient perceptions of quality of care at PHCs. Prior research has also established that increased perceptions of quality of care lead to increased demand, as demonstrated in both India[79] and in other developing countries.[80-83]

In this context, our PHC simulation models can prove useful, as demonstrated by the sensitivity analyses and configuration optimisation experiments in Section 4.3. For example, as the sensitivity analyses with consultation times for doctors closer to international levels showed, PHC resources become stressed even when only 30% of current healthcare demand is addressed at a PHC. Further, we also find that a significant proportion of childbirth patients (approximately 16-28% when the number of childbirth cases/day varies from 1-2) wait longer than two hours before receiving admission into the childbirth facility (bed) at a PHC. In response to these operational issues at higher demand levels, we have also demonstrated how the PHC models can be used to evaluate strategies for reconfiguring PHC resources to address the demand effectively prior to actually implementing them, as shown in Sections 4.3.1. - 4.3.3. This ties into the Indian government's programme of upgrading PHCs into HWCs and establishing an additional 150,000 HWCs. The findings from this study may be useful in specifying medical personnel numbers or childbirth room capacities (e.g., convert a few inpatient beds to childbirth beds) at these new/upgraded facilities. For example, if quality of care is to be increased in the upgraded PHCs or new HWCs by establishing guidelines regarding consultation durations, then the capacity of these individual facilities may also need to be expanded to accommodate both existing levels of demand and the increased levels of demand that may be experienced if quality of care increases.

From a more general methodological standpoint, the key research contribution of this study involves providing a proof-of-concept for modelling primary healthcare delivery units that are part of large hierarchical public health systems operating in underserved settings. In particular, our approach towards capturing the operational diversity of PHCs by applying a generic modelling and reconfigurable simulation approach to capture key operational characteristics could provide researchers studying other hierarchical health systems with a template towards modelling primary healthcare delivery. This could also assist in developing simulations of the network of PHCs in a given region. Simulations of the network of PHCs in a region can be used for many types of operational analyses associated with policy changes in healthcare administration in the region. For example, Fatma and Ramamohan[61] utilise the PHC models presented in this paper to develop a simulation of the network of PHCs in the district under consideration. They then demonstrate how the issue of increased wait times before admission to the childbirth facility in a PHC can be alleviated by diverting patients based on real-time predictions of their wait times generated at the time childbirth patients arrive at the PHC seeking admission. Similarly, in another working paper, we develop a simulation of the network of PHCs, CHCs, a DH and a makeshift COVID-19 care centre to determine how the public health system responds to COVID-19 caseloads under a specific pandemic response strategy. In the pandemic response strategy that we simulate, developed in collaboration with a clinical expert, the PHCs serve as testing and triaging centres for symptomatic patients with suspected COVID-19, wherein they are advised home isolation or hospitalisation depending upon the severity of their illness after diagnosis and triage. We adapt the PHC models that we present here to include testing and triaging pathways for suspected COVID-19 patients. Note that this operationalisation of PHCs as COVID-19 testing and triage centres is consistent with our recent visits to primary urban health centres (the equivalent of PHCs in urban metropolitan areas), and with the experiences of clinicians with expertise in COVID-19 management regarding the role PHCs are playing in the pandemic response strategy of the public health system.

The above studies illustrate the reusable nature of the generic PHC model that we have developed – specifically, they represent full model adaptation and reuse in the same setting, but for a different purpose. In contrast, the creation of simulation models of configurations 2 and 3 represent full model reuse in the same setting and for the same purpose (analysis of PHC operations). For example, in Fatma and Ramamohan[61], a real-time delay prediction and diversion module was added to the PHC models that we present here, and in the latter study described above, COVID-

19 testing and triaging pathways were added. These modified individual PHC models were then integrated into the network of public health facilities within the district.

Our simulation models also contribute to the whole healthcare facility simulation literature, with our search of the literature not yielding any other study that considered a primary care facility that serves outpatients, inpatients, childbirth cases and antenatal care patients. Finally, as described in 0, we introduce simple approximations for the estimation of the utilisation of a server (the PHC doctor in our model) with multiple job classes with significantly different exponential interarrival times and/or general service times, and also derive straightforward conditions under which the approximation is likely to hold. In particular, our proposed approach towards the conversion of the queueing system represented by the PHC doctor to an M/G/1 system makes its analysis significantly more tractable, in terms of analytical estimation of average time spent in the system, waiting time and average number of patients in the queue.

A challenge in developing such simulation models in the Indian context is obtaining adequate access to the facilities under consideration for a sufficiently long period of time to collect data required to fit distributions for every input parameter of the simulation. For instance, given the limited data maintained for inpatient length of stays, we were unable to observe inpatient admissions long enough to collect sufficient data to find the best-fitting distribution for inpatient length of stay. In such cases, we estimated these parameters based on our discussions with key medical personnel. We anticipate that the model will have to be updated when data for these parameters will become available. Further, we note that we have only included resources and operations associated with provision of medical care, and hence have not included maintenance/cleaning personnel, etc.

Overall, our work establishes the computational infrastructure required to analyse the operational capacity and performance of PHCs, and we anticipate that other researchers, policymakers and other stakeholders in health capacity planning will be able to utilise and/or adapt our simulation models to analyse PHC operations in their contexts.

# Appendices

## Appendix A.

### Overview of Public Health System in India

Public healthcare delivery in India is provided at four levels in the district (in increasing order of extent of services provided): a) the subcentre (SC), b) the primary health centre (PHC), c) the community health centre CHC), and d) the district hospital (DH). The primary healthcare infrastructure in India is designed as a three-tier system. Three tiers are SC at the base, PHC in the middle, and CHC at the top. The SC is the first contact point between the primary care system and community; it covers a population of 5,000 persons, and is limited to a coverage of 3,000 persons in hilly or tribal areas. An SC is manned by at least one auxiliary nurse midwife and provides maternal and childcare services, nutritional care, and immunisation among other services intended to improve population health. PHCs are small hospitals with one or two medical doctors who serve as the first point of contact between society and healthcare provided by formally trained doctors. CHCs were established to provide both primary and secondary care to the community. It was envisaged that people who require specialised care could access a CHC directly or if required, by referral from a PHC. A CHC is mandated to be a thirty-bedded hospital with four specialised doctors - in surgery, medicine, gynecology, and pediatrics - and supported by 21 paramedical and other staff. It acts as a referral unit for four or more PHCs. DHs are bigger hospitals in comparison to PHCs and CHCs, and were established to provide comprehensive secondary care and limited tertiary care. Per operational guidelines for DHs, each district is mandated to have a DH with the number of beds in the hospital ranging from 75 - 500, based on the population size and the geography of the district. Services provided at the DH are categorised as essential (general medicine, general surgery, ophthalmology, intensive care units, and radiology), desirable (dermatology, radiotherapy, dialysis service, etc.), and superspFfecialties (such as neurosurgery).

## Appendix B.

### Estimation of Service Time Parameters

For determining the distribution of the doctor's consultation time with outpatients, 60 observations were recorded across 6 PHCs per Table 2, and two outlier service time values were identified and removed from this dataset. Figure B1a depicts the histogram for the data. We conducted the Anderson-Darling (AD) normality test using the Minitab software and observed a p-value of 0.466 and an AD statistic of 0.348. Histogram plots are shown in Figure B1b and Figure B1c, respectively, for similar data collected for the laboratory and the pharmacy service times. With regard to the normality test for the laboratory service time, a p-value of 0.265 and an AD statistic of 0.465 was reported. From the 60 observations for the pharmacy service time recorded in the PHCs we visited, three outlier values were removed, and then the AD normality test was conducted which yielded a p-value of 0.327 and an AD statistic of 0.413.

With regard to negative values from the estimated normal distribution for the doctor's consultation time for outpatients, we truncate normal distribution at 30 seconds, the lowest consultation time observed during our data collection process. Similarly, the distributions of the pharmacy and laboratory service times are truncated at 40 seconds and 120 seconds, respectively, both of which are approximately equal to the lowest service times observed during the data collection exercise.

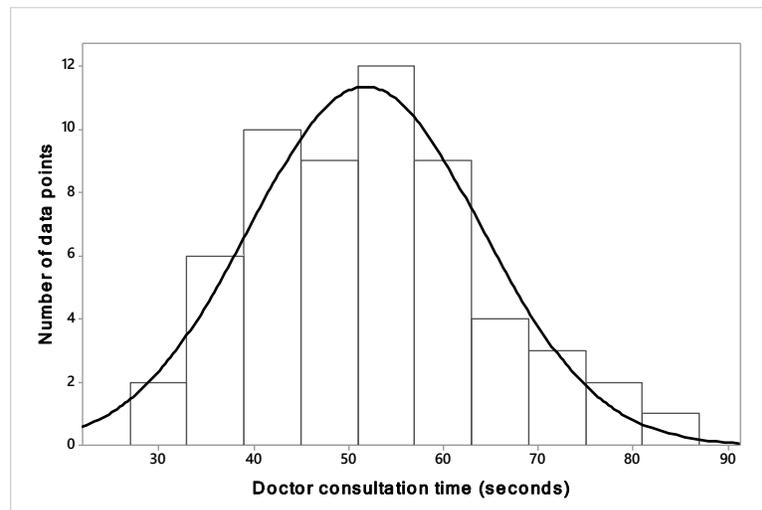

Figure B1a. Histogram of the doctor's consultation time data for outpatients

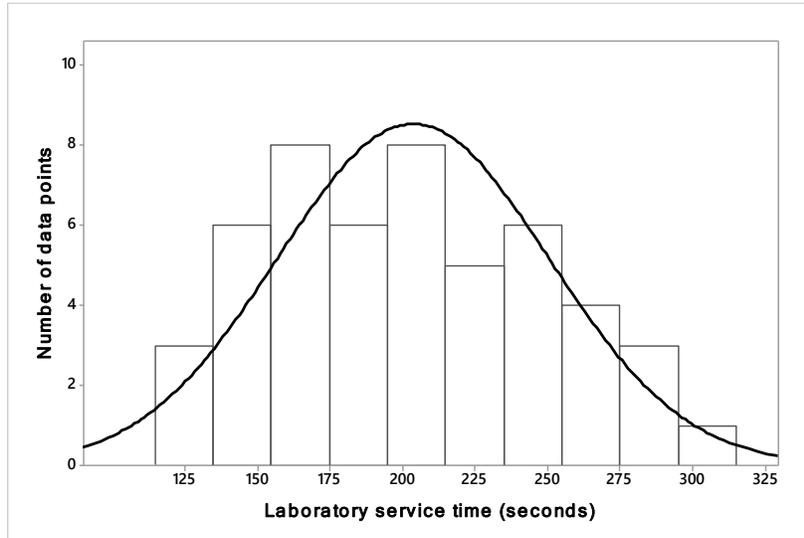

Figure B1b. Histogram of the laboratory service time data

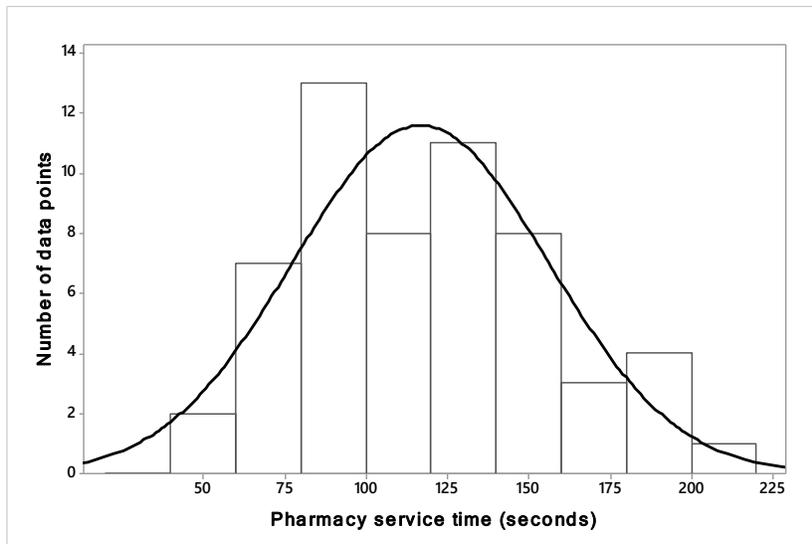

Figure B1c. Histogram of the pharmacy service time data

# Appendix C.

## Analytical Approximations for the Doctor's Utilisation

In this section, we utilise the notation defined in Section 4.1. First, we investigate the extent to which $\rho_o$ approximates $\rho_d$, given that outpatient arrival rates are nearly three orders of magnitude larger than inpatient and childbirth patient arrival rates. We derived the result below using the notion of the 'domination factor', the extent to which one job type dominates the other job types in terms of average arrival rates, service rates or utilisation in general. For example, the domination factor for outpatients in our queueing system can be expressed as follows: $d_o = \frac{\rho_o}{\rho_o + \rho_i + \rho_c}$. The domination factor may also represent a belief regarding the extent to which one job type dominates other job types, where data for precisely estimating each term in the above equation may not be available.

*Theorem C.1.* Consider a queuing system with a single server and $n$ types of jobs, with Poisson arrivals (with average arrival rates $\lambda_1, \lambda_2, \lambda_3, \ldots, \lambda_n$) and general service times for each job type (with corresponding average service rates $\mu_1, \mu_2, \mu_3, \ldots, \mu_n$). Let the utilisation of the server be a random variable with a symmetric and unimodal distribution $f_d$ with expected value $\rho_d$ and standard deviation $\sigma_d$, which may be estimated by a simulation in steady state that yields a single estimate of utilisation in each replication. Let the estimated (from the simulation) values of $\rho_d$ and $\sigma_d$ be $\hat{\rho}_d$ and standard deviation $\hat{s}_d$. Further, let $\rho_1, \rho_2, \rho_3, \ldots, \rho_n$ represent the average utilisations of the server, with $\rho_i = \frac{\lambda_i}{\mu_i}$ ($i = 1 - n$), if only a single type of job was considered for the system. Without loss of generality, let the first job ($i = 1$) be the dominant job type. Then $\rho_d$ can be approximated by $\rho_1$ at an $\alpha$ level of significance if $d_1 = \frac{\rho_1}{\sum_{\{i=1\}}^{n} \rho_i} > \frac{\hat{\rho}_d - k_\alpha \hat{s}_d}{\hat{\rho}_d}$.

*Proof.* Let the average utilisation estimated from the simulation be $\hat{\rho}_d$ (we assume one replication of the simulation yields a single steady state estimate of utilisation). Let the standard deviation of the utilisation be denoted as $\hat{s}_d$. Then if $f_d$ represents the distribution of the doctor's utilisation, we say that $\rho_1$ approximates $\rho_d$ with an $\alpha$ level of significance if $\rho_1 \in I$, where $I = (f_d^{-1}(\frac{\alpha}{2}), f_d^{-1}(1 - \frac{\alpha}{2}))$.

We now derive the conditions under which $\rho_1 \in I$.

Let $f_d^{-1}\left(\frac{\alpha}{2}\right) = \hat{\rho}_d - k_{\frac{\alpha}{2}} \hat{s}_d$ and $f_d^{-1}\left(1 - \frac{\alpha}{2}\right) = \hat{\rho}_d + k_{\left(1 - \frac{\alpha}{2}\right)} \hat{s}_d$. If $f_d$ is symmetric and unimodal, then $k_{\frac{\alpha}{2}} = k_{\left(1 - \frac{\alpha}{2}\right)} = k_\alpha$. We make the simplifying assumption that $f_d$ is symmetric and unimodal for the remainder of our analysis. Thus, the problem reduces to deriving the condition under which $\rho_1 \in I$, where $I = (\hat{\rho}_d - k_\alpha \hat{s}_d, \hat{\rho}_d + k_\alpha \hat{s}_d)$.

This is possible only if $|\hat{\rho}_d - \rho_1| < k_\alpha \hat{s}_d$.

Now, $\rho_d > \rho_1$ and therefore it is reasonable to assume that in steady state $\hat{\rho}_d > \rho_1$ (more details in subsequent section). Therefore we write $|\hat{\rho}_d - \rho_1| = \hat{\rho}_d - \rho_1$.

Therefore, $\rho_1 \in I$ if $\hat{\rho}_d - \rho_1 < k_\alpha \hat{s}_d$; that is, if $\rho_1 > \hat{\rho}_d - k_\alpha \hat{s}_d$.

Now $\rho_1 = d_1 \sum_{\{i=1\}}^n \rho_i = d_1 \rho_a$ and therefore $\hat{\rho}_d - d_1 \rho_a < k_\alpha \hat{s}_d$.

This implies that $\rho_1 \in I$ if $d_1 > \frac{\hat{\rho}_d - k_\alpha \hat{s}_d}{\sum_{\{i=1\}}^n \rho_i}$. Now, we can assume that $|\rho_a - \hat{\rho}_d| \approx 0$ (this is seen in Table C.1 below, and is also based on the analytical property of the queueing system under consideration that $\rho_a$ is the best analytical estimator of $\rho_d$, whereas $\hat{\rho}_d$ can be considered to be the best empirical estimator of $\rho_d$), and hence the above result can be written as $\rho_1 \in I$ if $d_1 > \frac{\hat{\rho}_d - k_\alpha \hat{s}_d}{\hat{\rho}_d}$.

The above approximation may be useful in situations where reasonably accurate arrival and service data is available for the dominant job type, but similar data is not available for the less frequent job types. This is applicable to the service system corresponding to the doctor in the PHC, where primary data is available for the doctor's consultation time for outpatients, whereas only point estimates (without uncertainty information) based on discussions with the medical staff are available for the arrival and service rates associated with inpatients and childbirth patients. In such a situation, our approximations can be used in the following manner: if there is reason to believe that one job type dominates other job types by a certain extent - for instance, its arrival rate is such that between 85 - 90% of jobs are contributed by this job type, and service rates for all jobs are approximately the same, then the utilisation of this job type lies within $k_\alpha \sigma_d$ of $\rho_d$ with probability $1 - \alpha$ if the condition in the above theorem is satisfied. Here $\alpha$ can be chosen such that $k_\alpha \sigma_d$ represents the desired maximum allowable deviation (e.g., 5%) from $\rho_d$. Satisfying the condition described by Theorem C.1 can just involve checking whether $d_1 > 1 - \frac{k_\alpha \hat{s}_d}{\hat{\rho}_d}$. Thus, for the above example, if the domination factor $d_1$ is believed (or estimated) to be between 85-90%, and the server's utilisation is required to be approximated with a maximum of 5% error, then $\frac{k_\alpha \hat{s}_d}{\hat{\rho}_d} = 0.05$, and thus in this case, the approximation cannot be used for any value of $d_1$ in the above range ($d_1 = 85 - 90\%$). The results in Table C.1 reflect this. For three configurations (1, 2 and the benchmark case), because the inpatient and childbirth service times are significantly higher (one and two orders of magnitude higher than outpatient service times), $d_o < 0.95$ for these three configurations, and hence by Theorem C.1, $\rho_o$ cannot be used to approximate $\rho_d$ with probability $1 - \alpha$, and the results in Table C.1 verify this. However, we note that even in these cases, the difference between $\rho_o$ and $\hat{\rho}_d$ is at maximum approximately 13%. For configuration 3, because childbirth services are not offered, $d_o > 0.95$, and hence, by Theorem C.1 (supported by the numerical evidence), it approximates $\rho_d$ with probability $1 - \alpha$.

We also explored the conversion of this system to an M/G/1 system by treating the less frequently arriving patient types as nonpreemptive 'setup' jobs, following the approach indicated in Hopp and Spearman.[84] This approach yields another approximation of $\rho_d$ (denoted by $\rho_{ap}$) via the conversion of this system to an M/G/1 queueing system, but also yields estimates for average outpatient waiting time, time spent in the system, etc. We now describe in detail how this conversion is achieved.

We describe the analysis of the server and the dominant job type in this queueing system by converting it to an M/G/1 system, which is a significantly simpler system to analyse than the queueing system with multiple types of jobs with nonpreemptive priority. This approach may be useful when the server and one particular job type (ideally the dominant job type) is the focus of the analysis, because the simplification of this system comes at the cost of information regarding waiting times and time spent in the system for the other job types.

We first consider the case when only one other patient type other than outpatients are served by the doctor. We achieve the conversion to an M/G/1 system by applying the approach provided in Hopp and Spearman[84] for calculation of effective process time of a machine when setups need to be performed between jobs. In our system, the "jobs" represent the dominant job type and the "setups" represent all other job types. Let $\lambda_1$ denote the average arrival rate of the dominant job type and $\mu_1$ represent its service rate. Therefore, we define average utilisation as $\rho_1 = \frac{\lambda_1}{\mu_1}$. Arrival of other job types can be thought of as the arrival of setups that can be attended to immediately after the current job (e.g., the dominant job type) is processed. If the rate of arrival of other job types (setups) is denoted by $\lambda_i$ ($i = 2 - n$), we can calculate the average number of dominant jobs after which a setup arrives. We denote this by $N_i$. Then, following the analysis presented in Hopp and Spearman, the effective average process time of the doctor for outpatients, including inpatients (setups), becomes:

$\frac{1}{\mu'_1} = \frac{1}{\mu_1} + \frac{1}{\mu_i N_i}$, where $\frac{1}{\mu_i}$ is the mean service time for setup $i$.

Therefore, it is clear that $\mu'_1 < \mu_1$, and hence $\rho'_1 > \rho_1$. It is then reasonable to assume that if a large number of replicate observations of the doctor's utilisation are obtained under steady state simulation conditions, $\hat{\rho}_d$ will also be greater than $\rho_1$.

Note that $\rho'_1$ can be further modified by considering the arrival of next job type as another type of setup, and thus the impact of all other non-dominant job types can be incorporated into the values of $\mu'_1$ and $\rho'_1$. Let the average utilisation of the server of such an M/G/1 system be denoted by $\rho_{ap}$. Note that $\rho_{ap}$ takes the other patient types into account, and hence is likely to be a better approximation of $\rho_d$ than $\rho_o$. We derive the following condition that $\rho_{ap}$ must satisfy to be a valid approximation of $\rho_d$.

*Theorem C.2.* Let $\rho_{ap}$ be an approximation of $\rho_d$ in the queueing system described In *Theorem C.1*, and define $r = \frac{k_\alpha \hat{s}_d}{\hat{\rho}_d}$. Then $\rho_{ap}$ approximates $\rho_d$ at an $\alpha$ level of significance if $d_1 \in (\frac{(1-r)\rho_1}{\rho_{ap}}, \min\{\frac{(1+r)\rho_1}{\rho_{ap}}, 1\})$, where $d_1 = \frac{\rho_1}{\sum_{\{i=1\}}^n \rho_i}$.

*Proof.* Let $\rho_{ap}$ be any approximation of $\rho_d$ that takes into account the impact of non-dominant job types. $\rho_{ap}$ approximates $\rho_d$ at an $\alpha$ level of significance if $|\hat{\rho}_d - \rho_{ap}| < k_\alpha \hat{s}_d$. We have $r = \frac{k_\alpha \hat{s}_d}{\hat{\rho}_d}$, therefore $\rho_{ap}$ approximates $\rho_d$ at an $\alpha$ level of significance if $\rho_{ap} \in (\hat{\rho}_d - r\hat{\rho}_d, \hat{\rho}_d + r\hat{\rho}_d)$. Now, given that $\sum_{\{i=1\}}^n \rho_i$ is the standard analytical estimator of $\rho_d$, and $d_1 = \frac{\rho_1}{\sum_{\{i=1\}}^n \rho_i}$, we can replace $\hat{\rho}_d$ with $\rho_1/d_1$ above. Rearranging terms completes the proof.

We suggest that the above approximation can also be used when the focus of interest is the server's utilisation and outcomes related to the dominant job type (e.g., average dominant job type waiting time, time spent in the system). This may be useful in situations where an analyst may have limited access to literature regarding queueing systems with jobs of different priorities, and in this case, the more common Kingman approximations[85] for the average wait time, length of stay and number of entities in an M/G/1 queueing system may be used. The above representation of the system can be used in situations where, in a manner similar to the case when $\rho_o$ approximates $\rho_d$, $d_1$ and $\rho_1$ are known with a high degree of accuracy, and $\lambda_i$ and $\mu_i$ for the remaining jobs are known with lower accuracy. For example, the M/G/1 approximation may be applied in a situation with $n$ job types, if it is known that: a) non-dominant job types arrive on average after $k$ dominant jobs are processed, and b) detailed information (e.g., primary or secondary data) regarding service times for these job types is not available, and it is only known that on average they require a certain fraction of the service time of the dominant job type.

The numerical results in Table C.1 suggest that, as expected, $\rho_{ap}$ is a significantly better approximator of $\rho_d$ than $\rho_o$. Even for for configuration 2, wherein the condition in Theorem C.2 is not satisfied by $d_o$, we see that the difference between $\rho_{ap}$ and $\hat{\rho}_d$ is approximately 9.4%, lower than the corresponding Table C.1 maximum difference of approximately 13% for $\rho_o$.

Table C.1. Simulation-based validation of analytical approximations of doctor's utilisation.

| PHC Configuration | $\hat{\rho}_d$ | $\rho_o$ (p-value, % difference from $\hat{\rho}_d$) | $\rho_{ap}$ (p-value, % difference from $\hat{\rho}_d$) |
|---|---|---|---|
| Configuration 1 | 0.122 | 0.109 (0.004, 11.9) | 0.1129 (0.05, 7.4) |
| Configuration 2 | 0.109 | 0.0967 (0.006, 12.7) | 0.0988 (0.02, 9.4) |
| Configuration 3 | 0.099 | 0.0969 (0.39, 2.2) | 0.0973 (0.51, 1.8) |
| Benchmark configuration | 0.870 | 0.8334 (0.004, 4.4) | 0.865 (0.81, 0.6) |